\documentclass{aa}  

\usepackage{graphicx}
\usepackage{txfonts}
\usepackage{multirow}
\usepackage{array}
\usepackage{xcolor}
\usepackage{soul}

\newcommand\Tstrut{\rule{0pt}{2.8ex}}         
\newcommand\Bstrut{\rule[-1.1ex]{0pt}{0pt}}   

\begin{document}

   \title{The chemical DNA of the Magellanic Clouds}
   
   \subtitle{VI. Origin and evolution of neutron-capture elements in the SMC}

   \author{M. Palla\inst{1,2,3}
          \and
          A. Mucciarelli\inst{2,3} 
          \and 
          D. Romano\inst{3}
          \and
          S. Anoardo\inst{2}
          \and 
          F. Matteucci\inst{4}
          }

   \institute{
            INAF - Osservatorio Astrofisico di Arcetri, Largo E. Fermi 5, 50125, Firenze, Italy\\
            \email{marco.palla@inaf.it}
        \and    
            Dipartimento di Fisica e Astronomia “Augusto Righi”, Alma Mater Studiorum, Via Gobetti 93/2, I-40129 Bologna, Italy
        \and
            INAF, Osservatorio di Astrofisica e Scienza dello Spazio, Via Gobetti 93/3, I-40129 Bologna, Italy
        \and 
            Dipartimento di Fisica, Sezione di Astronomia, Università di Trieste, Via G. B. Tiepolo 11, 34143 Trieste, Italy
             }

   \date{Received ; accepted }

 
  \abstract
   {In the context of galactic archaeology, the study of the Small Magellanic Cloud (SMC) is of crucial importance, as it represents a unique opportunity to study a nearby massive dwarf system. However, theoretical studies of the chemical evolution of this galaxy are strikingly 
   lacking.}
   {In this study, we investigate the chemical enrichment of the SMC galaxy. Besides $\alpha$ and Fe-peak elements, we devote particular attention to the evolution of neutron-capture elements with different origin, namely r-process (Eu), weak s-process (Zr) and main s-process (Ba, La).}
   {We develop chemical evolution models that use as input the star formation histories obtained from colour-magnitude diagram fitting. We follow in detail the chemical feedback provided by a large variety of nucleosynthetic sources. Model predictions are compared with recent abundance measurements for the SMC.}
   {The developed framework reproduces well all the observables for elements up to the Fe-peak. 
   The abundance patterns of n-capture elements are simultaneously reproduced only by assuming an enhanced contribution from the delayed r-process at low metallicity and a top-lighter IMF relative to the reference IMF by \citet{Kroupa01}. In this way, both the observed very high plateau in [Eu/Fe] and the rising trends in [s-process/Fe] ratios can be reproduced by the models.}
   {This study provides for the first time information on the evolution of several n-capture elements in a massive dwarf irregular galaxy, also providing insight on several ingredients driving galactic evolution. 
   Moreover, this work provides a test-bed for further modelling of the SMC in the context of the numerous surveys  that will target the Magellanic Clouds in the next years.}


   \maketitle
%

\defcitealias{Mucciarelli23}{Paper I}
\defcitealias{Mucciarelli23_GCs}{Paper II}
\defcitealias{Anoardo25}{Paper IV}

\section{Introduction}
\label{s:intro}

The Milky Way (MW) most massive satellites, the Large and Small Magellanic Clouds (LMC and SMC, respectively), offer the unique opportunity to study massive dwarf systems at very small distances ($\sim$50 kpc and $\sim$62.5 kpc from the Sun, \citealt{Pietrzynski19,Graczyk20}) before the Galaxy cannibalisation.
The SMC, in particular, has a total mass of 2 $\times$ $10^9$ M$_\odot$ \citep{Stanimirovic04}. This is about one order of magnitude lower than the mass of the LMC. The stellar mass of the SMC is 5-7$\times 10^8$ M$_\odot$ \citep{Rubele18,Massana22}, which is comparable to the mass of the main merger of the MW, namely Gaia-Sausage-Enceladus (e.g. \citealt{Helmi18}).

In the last decade, there has been a renewed interest in the evolution of the Magellanic Clouds and the SMC in particular, as evidenced by several dedicated photometric surveys, such as VMC (VISTA survey of the Magellanic Clouds system, \citealt{Cioni11}), STEP (the SMC in Time: Evolution of a Prototype interacting late-type dwarf galaxy, \citealt{Ripepi14}), SMASH (Survey of the MAgellanic Stellar History, \citealt{Nidever14}), and VISCACHA (VIsible Soar photometry of star Clusters in tApii and Coxi HuguA, \citealt{Maia19}). 
Chemical analyses of high-resolution spectra of statistically significant samples of SMC red giant branch (RGB) stars, instead, have only been presented recently. The APOGEE-2 survey measured abundances for $\sim$1000 SMC RGB stars, in particular Fe, $\alpha$-elements \citep{Nidever20}, Al, Ni, Ce \citep{Hasselquist21}, and Mn \citep{Fernandes23}. 
On the other hand, \citet{Mucciarelli23} presented FLAMES-GIRAFFE abundances of 14 elements (from O to La) for around $\sim 200$ SMC stars, now complemented by \citet{Anoardo25}, where Eu abundances are presented for a comparable number of objects in the SMC.

Despite recent observations provide an almost complete chemical inventory for the galaxy, models of galactic chemical evolution (GCE) focusing on the SMC are lacking. 
After the pioneering works by \citet{Pagel98} and \citet{Tsujimoto09} on the reproduction of the age-metallicity relation and stellar metallicity distribution observed for SMC Globular Clusters (GCs) and low-resolution observations of SMC field stars, the only exception is presented in \citet[][see their Section 5]{Hasselquist21}. However, in their work, models were built exclusively to fit the observed trends either for [Si/Fe] or [Mg/Fe], without addressing in detail the other available constraints for the galaxy. 
In this paper, we develop more detailed chemical evolution models for the SMC. The study of environments that evolved differently from the MW is important to resolve uncertainties, degeneracies, and questions still present when dealing with Galactic studies \citep[see, e.g.,][for a review]{Matteucci21}. 
Among them, a striking example is represented by elements beyond the Fe-peak (A$\gtrsim 60$) that are produced primarily by neutron (n)-capture slow (s-) and rapid (r-) processes. Indeed, the understanding of their astrophysical production sites has become one of the major topics in stellar physics and GCE (see e.g. \citealt{Prantzos20,Koba20all,Molero2023MNRAS.523.2974M} and references therein).

On the other hand, Local Group (LG) MW satellites will be privileged targets for forthcoming multi-object spectrographs, such as 4MOST (4-m Multi-Object Spectrograph Telescope, see \citealt{4MOSTpaper}) and MOONS (Multi-Object Optical and Near-infrared Spectrograph, see \citealt{Cirasuolo20}), with the Magellanic Clouds in particular as main targets of several subsurveys (e.g. \citealt{Cioni19,Gonzalez20}).
Therefore, establishing a solid modelling framework for such systems is vital in prevision of future surveys, in order to provide a useful reference and test-bed for the data revolution ($>$ one order of magnitude increase in the data amount, see e.g. \citealt{Gonzalez20}) taking place in the next years.\\

In light of this, in the present work we aim at developing detailed models of chemical evolution to reproduce the SMC galaxy.
After a proper testing of the theoretical framework, based on star formation histories (SFHs) obtained by means of colour-magnitude diagrams (CMDs) from photometric surveys \citep{Rubele18,Massana22}, we focus on the evolution of n-capture elements in the SMC. 
To do so, we exploit the data from \citetalias{Mucciarelli23} (hereafter \citetalias{Mucciarelli23}) and \citetalias{Anoardo25} (hereafter \citetalias{Anoardo25}), where SMC abundances for Zr, Ba, La and Eu are available. This means testing elements produced via different production channels, namely weak s-process (Zr), main s-process (Ba and La) and r-process (Eu).
In turn, this allows us to perform one of the first thorough theoretical studies on n-capture elements evolution in an environment different from that of the Galaxy. Indeed, despite the presence of several programmes targeting heavy elements in the LG (e.g. \citealt{Letarte10,Hill19,Reichert20} and references therein), works on the theoretical interpretation of n-capture element abundances in galaxies either are missing or focus only on fewer elements (Ba and Eu at most, see e.g. \citealt{Hirai2015,Vincenzo15,Molero21,Palla25}).\\

The paper is organised as follows. 
In Section \ref{s:model}, we present the framework adopted to model the chemical evolution of the SMC. 
In Section \ref{s:data} we provide a brief overview of the datasets adopted for comparison 
with the model outputs. 
In Section \ref{s:results} we compare the model predictions to the observations, focusing in Section \ref{s:ncapture_results} on the study of n-capture elements. 
Finally, in Section \ref{s:discussion_conclusion} we discuss our findings and draw our main conclusions.

\section{A new chemical evolution framework for the SMC}
\label{s:model}

In this Section, we present the chemical evolution framework adopted for the SMC. 
At variance with what commonly done in the literature, where parametric forms for the gas inflows/outflows are imposed and the SFH is obtained by assuming a star formation efficiency (SFE) to convert the available gas in stars,
here we consider SFHs obtained through CMD fitting \citep{Rubele18,Massana22} and, by assuming a SFE for each timestep, we let the gas budget evolve accordingly to the input star formation rate (SFR). 

The latter approach has the advantage to capture variations of the SFR over small timescales, as well as tracing multiple peaks in SF and gas accretion, without implementing additional fine-tuned parameters (e.g. onset times of different SF bursts and/or gas accretion episodes, e.g. \citealt{DeVis21,Palla24}). 
In this way, it allows to test more easily the different derived SFHs (see \ref{ss:sfh}) against other galactic observables, namely the age-metallicity relation, the stellar metallicity distribution function (MDF) and the [X/Fe] vs. [Fe/H] patterns.

\subsection{Star formation histories}
\label{ss:sfh}

\defcitealias{Rubele18}{R18}
\defcitealias{Massana22}{M22}

In this work, we adopt the SFHs inferred from SMC CMDs by \citet{Rubele18} and \citet{Massana22}. 
Going into the details of the derivation of the two SFHs is beyond the scope of this work. Below, we provide both a brief description.

In \citet[hereafter \citetalias{Rubele18}]{Rubele18}, the SFH across the entire main body and wing of the SMC is recovered using tile images from the VMC survey (\citealt{Cioni11}) in the $Y\ J\ Ks$ filters. The analysis was carried out in 168 subregions, applying a CMD reconstruction method that returns the best-fitting SFR($t$) (with age–metallicity relation, distance, mean reddening, see \citealt{Rubele15}). 
In \citet[hereafter \citetalias{Massana22}]{Massana22}, instead, the SFH of the SMC is obtained by using deep $u\ g\ r\ i\ z$ photometry of the second data release of the SMASH survey (\citealt{Nidever21}). Here, the analysis was performed on 74 regions with a similar number of stars using Voronoi binning \citep{Cappellari03} and applying a CMD reconstruction method for the SFH of the galaxy.

\begin{figure}
    \centering
    \includegraphics[width=0.99\linewidth]{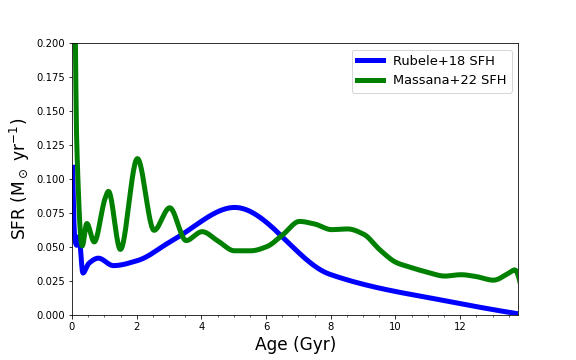}
    \caption{Global SFHs for the SMC galaxy as derived by \citet[][blue line]{Rubele18} and \citet[][green line]{Massana22}.}
    \label{fig:SFHs}
\end{figure}

As in our work we focus on the SMC integrated properties, for both studies we use integrated SFHs by summing up the contribution of each bin/subregion at each 
timestep. 
Moreover, it is worth noting that, as in \citetalias{Rubele18} the SFH is given in the form of a step function, we smooth it\footnote{We adopt a Piecewise Cubic Hermite Interpolating Polynomial (PCHIP) interpolation. This choice is due to the uneven (log spaced) sampling of the \citetalias{Rubele18} SFH that is subject to overfitting when other interpolations are applied \citep{Fristch84}.} to obtain a more realistic behaviour of the SFR with time. This procedure is not necessary for \citetalias{Massana22}, as they already provided a smoothed final solution for their SFH \citep[see, e.g.,][for details]{Ruiz18}.\\

The resulting SFHs as functions of age/lookback time are shown in Fig. \ref{fig:SFHs}.
They show a quite different evolution, with \citetalias{Rubele18} having a smoother behaviour relative to \citetalias{Massana22}, which displays much more prominent SF peaks, especially at recent times (age~$\lesssim2-3$ Gyr). 
This marked increase in SF around 3 Gyr ago in \citetalias{Massana22} was also observed in previous analyses (e.g. \citealt{Harris04}) and is associated with interactions/close encounters with the LMC. These would also trigger the present-time SF burst observed by both \citetalias{Rubele18} and \citetalias{Massana22}.
On the other hand, the larger SFR at older ages in \citetalias{Massana22} is attributed by the authors to the magnitude limit of the VMC survey, as the 10 Gyr old main-sequence turnoff lies very close to the 50\% VMC completeness limit.
The differences in the SFHs inferred by \citetalias{Rubele18} and \citetalias{Massana22} are also reflected in the stellar masses integrated over the galaxy lifetime ($5.2 \cdot10^{8}\ {\rm M_\odot}$ for \citetalias{Rubele18}, $6.7 \cdot10^{8}\ {\rm M_\odot}$ for \citetalias{Massana22}), whereas the timescales for stellar mass accumulation are similar (50\% of the stellar mass formed in the first $\simeq8.5$ Gyr).

\subsection{Chemical evolution calculations}
\label{ss:chemevo_model}

To compute the chemical evolution of the galaxy under scrutiny, we rest on the basic equation that describes the evolution of a given element $i$ (see, e.g. \citealt{Matteucci21}):
\begin{equation}
    \dot{M}_i (t) = -\psi(t)\, X_i(t)\, + \,R_i(t)\, +\, \dot{M}_{i,flows}(t),
    \label{eq:basic_chemevo}
\end{equation}
The first term on the right-hand side of Eq. \eqref{eq:basic_chemevo} corresponds to the rate at which an element $i$ is removed from the ISM due to star formation, with the rate at a specific time $\psi(t)$ taken from the SFHs illustrated in Section \ref{ss:sfh}. 

The term $R_i$ (see, e.g. \citealt{Palla20b} for the complete expression) takes into account the nucleosynthesis from different stellar sources (see \ref{ss:nucleosynthesis} for details), weighted according to the initial mass function (IMF). The products originating from binary systems, such as Type Ia SNe and merging neutron stars, are included in $R_i$, properly accounting for the respective delay-time-distributions (see \citealt{Matteucci09,Palla25} for more details).
We adopt as a reference the IMF of \citet{Kroupa01} to be consistent with the assumptions made to derive the SMC SFHs; however, we also allow IMF variations (e.g. \citealt{Kroupa93}) to probe the impact of the latter on the galactic chemical evolution.

The last term of Eq. \eqref{eq:basic_chemevo} refers to the gas flows, namely inflows and outflows. 
These are computed in order to obtain the required gas mass for a given SFR $\psi(t)$ and SFE $\nu(t)$, with the latter a free parameter of the model variable over time, in this way:
\begin{equation}
    M_{gas}(t) = \psi(t) / \nu(t).
    \label{eq:gas_budget}    
\end{equation}
In practice, if $M_{gas}(t) > M_{gas}(t-\Delta t) + R_i(t-\Delta t)$, that is the gas mass at timestep $t$ is larger than the sum of the gas mass and the returned mass from dying stars at  timestep $t-\Delta t$, we impose that a gas amount equal to $M_{gas}(t) - ( M_{gas}(t-\Delta t) + R_i(t-\Delta t))$ is accreted by the galaxy. This gas is assumed to have a primordial chemical composition. 
Conversely, if $M_{gas}(t) < M_{gas}(t-\Delta t) + R_i(t-\Delta t)$, the gas excess $(M_{gas}(t-\Delta t) + R_i(t-\Delta t))-M_{gas}(t)$ is ejected from the system. This ejected gas has a chemical composition equal to that of the ISM at the timestep $t$.

\subsection{Nucleosynthesis prescriptions}
\label{ss:nucleosynthesis}

Among the basic ingredients of chemical evolution models are the stellar yields, namely the amounts of different chemical elements that stars produce and eject into the ISM at their deaths. We adopt grids of stellar yields for different types of stars well-tested on the MW data:
  
\begin{itemize}
    
    \item for low- and intermediate-mass stars (LIMS) we adopt the set of yields for non-rotating stars available on the web pages of the FRUITY\footnote{http://fruity.oa-teramo.inaf.it} data base \citep{Cristallo09,Cristallo11,Cristallo15};
    
    \item for massive stars, we adopt the yields by \citet{Nomoto13} that include elements up to the Fe-peak, assuming a hypernova (HN) fraction that varies with metallicity. In particular, we assume that the fraction of stars above 25 $M_\odot$ that die as HNe is 95\% up to [Fe/H]=$-$2.5 dex and then decreases to 0\% for [Fe/H]$>-$1 dex. This is in line with widespread assumptions of a variable HN fraction with metallicity (e.g. \citealt{KobaNaka11,Mucciarelli22}). 
    As \citet{Nomoto13} yields do not include those of n-capture elements, to account for the latter we adopt the yields by \citet{Limongi18}\footnote{we do not adopt \citet{Limongi18} for all the elements as for the problems by these yields in reproducing MW abundance patterns for several elements up to the Fe-peak \citep[e.g. Mg, see][]{Prantzos18,Palla22}.} 
    In particular, we set as reference the set R150 (namely, stars with uniform rotational velocity at $v_{rot}=$150 km s$^{-1}$); however, in the paper we test also different sets/velocity distributions; 
    
    \item for Type Ia SNe, we assume an equal mixture of near-$M_{Ch}$ and sub-$M_{Ch}$ progenitors. In fact, there are claims in the literature that both progenitors contribute to the chemical evolution of the Galaxy (see, e.g. \citealt{Koba20SNIa,Palla21}). The yields adopted for the two classes of progenitors are from \citet[for near-M$_{Ch}$]{Leung18} and \citet[for sub-M$_{Ch}$]{Leung20};
    
    \item to include the products of the r-process, we consider both magneto-rotational driven SNe (MRD-SNe) and merging of compact objects (MNS). For MRD-SNe we adopt the yields from model L0.75 of \citet{Nishimura17}, while for MNS we use the empirical yield measured by \citet{Watson19} for kilonova AT2017gfo \citep[see also][]{Molero2023MNRAS.523.2974M}.  
    At variance with the yields described in previous points, the latter prescriptions do not include dependences on the progenitor mass/metallicity: however, this is a common issue in modelling the chemical evolution of the r-process elements (see \citealt{Molero25,Palla25} and references therein).
    
\end{itemize}

\section{Observational data}
\label{s:data}

The spectroscopic data used in this work are taken from \citetalias{Mucciarelli23} and \citetalias{Anoardo25} for field stars and from \citet{Mucciarelli23_GCs} (hereafter \citetalias{Mucciarelli23_GCs}) for three GCs.

Concerning the SMC field, \citetalias{Mucciarelli23} analysed GIRAFFE-FLAMES@VLT spectra of 206 RGB stars, deriving chemical abundances for $\alpha$-elements (Mg and Si), iron-peak elements (Fe and Ni), and s-process elements (Zr, Ba and La). No r-process element abundances could be derived, as the FLAMES setups adopted in their study do not include any transition of r-process elements. 
On the other hand, \citetalias{Anoardo25} measured Eu abundances from GIRAFFE-FLAMES@VLT spectra of 209 RGB stars in the SMC (158 in common with the sample of \citetalias{Mucciarelli23}), utilizing a different instrument setup. We use the abundances of \citetalias{Anoardo25} to investigate the r-process enrichment in the SMC. Both samples cover a large metallicity range, $-2.3 \lesssim$[Fe/H]/dex$\lesssim -0.5$, with most targets having [Fe/H]$>-$1.4 dex.
Additionally, we include the abundances of three massive GCs in the SMC, namely NGC~121 \citep[the oldest SMC globular cluster, with an age of 10.5$\pm$0.5 Gyr,][]{glatt08a},  NGC~339 \citep[with an age of 6$\pm$0.5 Gyr,][]{glatt08b} and NGC~419 \citep[with an age of 1.4$\pm$0.2 Gyr,][]{glatt08b}, all of them analysed  using UVES-FLAMES@VLT \citepalias{Mucciarelli23_GCs}.
This is the only sample of chemical abundances for SMC clusters derived from high-resolution spectroscopy and it provides a time-resolved reconstruction of the chemical enrichment history of the SMC.

\section{Reproducing the main SMC chemical properties}
\label{s:results}

In the following, we present the predictions of the chemical evolution models described in Section \ref{s:model} and their comparison with the datasets illustrated in Section \ref{s:data}. 

The models adopting the two different SFHs have different star formation efficiencies $\nu$ calibrated in order to reproduce several observables in the galactic system, namely the stellar metallicity distribution function (MDF), the age-metallicity relation, and the [X/Fe] vs. [Fe/H] patterns of chemical elements X of well-known and distinct nucleosynthetic origin (Mg, Si, Ni).

\subsection{Metallicities}
\label{ss:PhysQuant_MDF_ageFeH}

\begin{figure}
    \centering
    \includegraphics[width=0.99\linewidth]{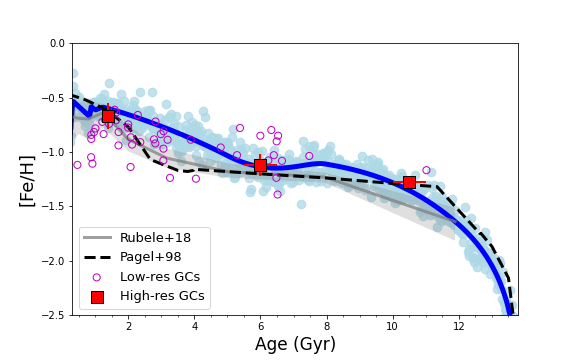}
    \includegraphics[width=0.99\linewidth]{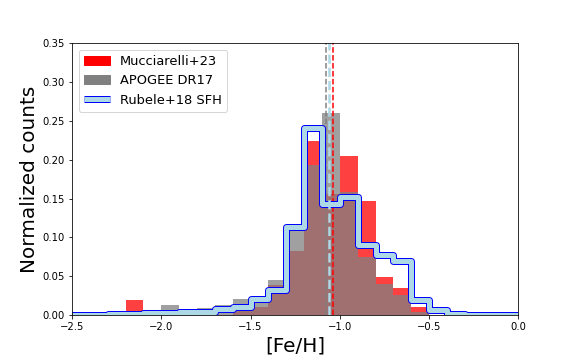}
    \caption{Age-metallicity relation (top panel) and MDF (bottom panel) as predicted by the reference chemical evolution model adopting \citet{Rubele18} SFH (blue line, circles and histogram, respectively). The predicted age-metallicity relation is compared with the CMD-extracted relation by \citet[][grey line with confidence intervals]{Rubele18}, the thereotical curve by \citet[][bursting model, black dashed line]{Pagel98}, SMC GCs from low-resolution \citep[][magenta empty circles]{Parisi09,Parisi15,Parisi22,Dias21} and high-resolution studies \citep[][red filled squares]{Mucciarelli23_GCs}. Observed MDFs are from \citet[][red histogram]{Mucciarelli23} and \citet[][APOGEE DR17, grey histogram]{Hasselquist21}.}
    \label{fig:AgeFeH_MDF_Rubele}
\end{figure}

The age-metallicity relation predicted by the model adopting the \citetalias{Rubele18} SFH is compared with the relevant data in Fig. \ref{fig:AgeFeH_MDF_Rubele}, top panel. 
For this model, we adopt a value of the free parameter $\nu=$0.015 Gyr$^{-1}$ up to an evolutionary time of $t=13$ Gyr, increasing it to $\nu=$0.05 Gyr$^{-1}$ in the last Gyr of evolution. 
The increase in the SFE can be explained in light of the large ($\times$ 3-4) and rapid increase in the SFR in the last Gyr of evolution found by \citetalias{Rubele18}, which can be interpreted as a starburst stimulated by the recent interaction with the LMC. 
Indeed, without increasing the SFE, larger SFRs would result in larger gas accretion, which in turn would lead to strong gas dilution and therefore lower metallicities than observed in the late galactic evolutionary phases.

In Fig. \ref{fig:AgeFeH_MDF_Rubele}, top panel, the blue solid line represents the age-metallicity relation predicted by our homogeneous GCE model. The light blue points are the predictions of a synthetic model that adds a random error, equal to the typical abundance uncertainties of the data samples described in Section \ref{s:data} ($\simeq 0.1$ dex), to the abundances of the synthetic stars formed at each time $t$ (see also \citealt{Palla22}). For each chemical element, we define a ‘new abundance’,
\begin{equation}
    [\rm X/H]_{new}(t) = [\rm X/H](t) + \mathcal{N}([\rm X/H], \sigma[\rm X/H]),
    \label{eq:synthetic_model}
\end{equation}
where $\mathcal{N}$ is a random function with normal distribution. 
In this way, we ensure a fair comparison between the model predictions and the observations, avoiding any artificial discrepancy caused by data uncertainties. 
Our model predictions align well with both the SMC GC data of \citetalias{Mucciarelli23_GCs} and the theoretical relation of \citet{Pagel98}. The only significant difference between the relation by \citet{Pagel98} and ours is found at ages comprised between 2 and 4 Gyrs, where our model predicts metallicities up to $\sim$0.4 dex higher. 
Unfortunately, no high-resolution data are available for GCs in this age range and previous low-resolution works \citep[][]{Parisi09,Parisi15,Parisi22,Dias21} show a very large metallicity spread. This spread in low-resolution data is also observed for younger GCs (ages $\sim$1 Gyr), with some of the clusters reaching [Fe/H] values lower than both the GC NGC 419 observed at high-resolution and model predictions by almost 0.5 dex. A possible explanation of the metallicity spread could reside in clumpy SF within the SMC late evolution, as it can be inferred from the particular galaxy morphology \citep[e.g.][]{Zartisky00,Martinez19}.

In Fig. \ref{fig:AgeFeH_MDF_Rubele}, bottom panel, we compare the predicted MDF with the observed ones from \citetalias{Mucciarelli23} and APOGEE DR17  \citep{Hasselquist21}. 
The model distribution peaks at [Fe/H] slightly below $-1$ dex, alike the observed distributions, showing also a very similar median ($<{\rm[Fe/H]}>_{model}=-1.06$ dex, $<{\rm[Fe/H]}>_{APOGEE}=-1.07$ dex, $<{\rm[Fe/H]}>_{M23}=-1.04$ dex, see the dashed lines in the figure).
The only visible discrepancy between observations and model outcome resides in the metal-rich tail of the MDF, where predictions show a slightly higher fraction of stars than both \citetalias{Mucciarelli23} and APOGEE distributions.\\

\begin{figure}
    \centering
    \includegraphics[width=0.99\linewidth]{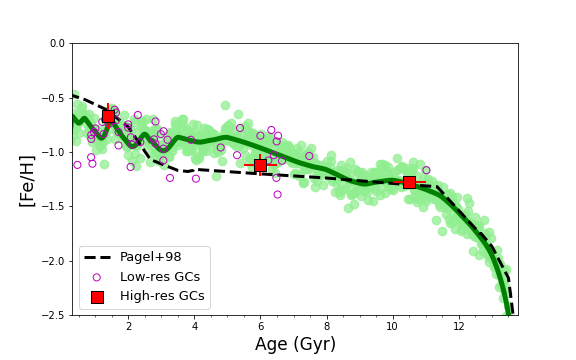}
    \includegraphics[width=0.99\linewidth]{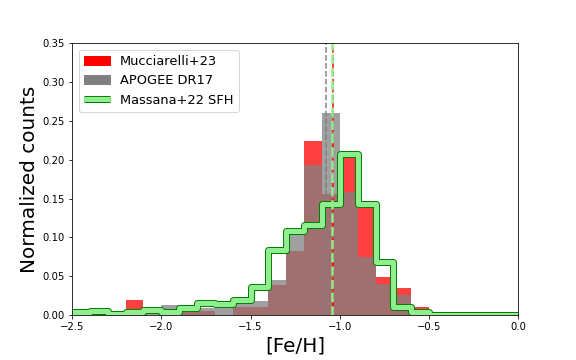}
    \caption{Age-metallicity relation (top panel) and MDF (bottom panel) as predicted by the reference chemical evolution model adopting \citet{Massana22} SFH (green curve, circles and histogram, respectively). Data as in Fig. \ref{fig:AgeFeH_MDF_Rubele}.}
    \label{fig:AgeFeH_MDF_Massana}
\end{figure}

The age-metallicity relation predicted by the model adopting the \citetalias{Massana22} SFH is compared with the relevant data in Fig. \ref{fig:AgeFeH_MDF_Massana}, top panel.
In this case, we adopt a SFE $\nu=$0.01 Gyr$^{-1}$ up to $t=11$ Gyr, increasing it to $\nu=$0.05 Gyr$^{-1}$ in coincidence with the multiple bursts of SF observed by \citet{Massana22} in the last 3 Gyrs. The adoption of a higher SFE allows us to reproduce the metallicity of the young SMC GC NGC 419 that otherwise would be underestimated, as a large increase in the average SFR would imply significant gas dilution if not accompanied by an increase in the star formation efficiency.
In general, we note that the model adopting \citetalias{Massana22} SFH produces metal-richer stars at intermediate ages (4-8 Gyr) relative to the model adopting \citetalias{Rubele18} SFH. However, the model with \citetalias{Massana22} SFH also results in lower metallicities at younger ages ($\lesssim$ 4 Gyr), which further highlights the effect of SF peaks in metallicity dilution as due to the injection of fresh gas into the galaxy ISM. 
Despite the differences relative to the \citetalias{Rubele18} SFH model, the general trend shown by the observations is still well matched, with the synthetic model adopting \citetalias{Massana22} reproducing the abundances of SMC GCs observed both at high- and low-resolution. 
The predicted MDF (green histogram in Fig. \ref{fig:AgeFeH_MDF_Massana}, bottom panel) reflects the differences highlighted in the age-metallicity relation: the distribution shows a peak skewed towards metal-rich stars, between $-0.9$ and $-1$ dex in [Fe/H], still in good agreement with APOGEE and \citetalias{Mucciarelli23} observations. At the same time, predictions do not show an excess of stars in the metal-rich tail ([Fe/H]$\gtrsim - 0.8$ dex) as for \citetalias{Rubele18}, but instead a slight overabundance in metal-poor stars ($-1.5 \lesssim$[Fe/H]/dex$\lesssim - 1.3$).
Nevertheless, the model distribution aligns well with the observations, with a median ($<{\rm[Fe/H]}>_{model}=-1.04$ dex) close to those of the data ($<{\rm[Fe/H]}>_{APOGEE}=-1.07$ dex, $<{\rm[Fe/H]}>_{M23}=-1.04$ dex). 

\subsection{$\alpha$ and Fe-peak elements abundance patterns}
\label{ss:XFe_FeH}

Another necessary step to test the models is to look at [X/Fe] vs. [Fe/H] diagrams for elements with well-known nucleosynthetic origin, for which we already have solid predictions in the context of the MW and the solar vicinity.
Building upon the sample of available elements in \citetalias{Mucciarelli23}, we look at Mg, Si and Ni. These three elements are synthesised by means of different processes and stellar sources / timescales: Mg is mainly produced by hydrostatic burning in massive stars \citep{WW95}, Si is made during explosive burning in massive stars with non-negligible contribution from Type Ia (\citealt{Romano10,Koba20all} and references therein), and Ni is primarily produced by Type Ia SNe (\citealt{Koba20SNIa,Palla21} and references therein). 
Moreover, the [X/Fe] vs. [Fe/H] patterns for Mg, Si and Ni observed by \citetalias{Mucciarelli23} are very similar to the ones in APOGEE \citep[][see Fig. \ref{fig:comparison_APOGEE} in Appendix]{Hasselquist21}, ruling out spurious patterns due to particular observational biases.\\

\begin{figure*}
    \centering
    \includegraphics[width=0.925\textwidth]{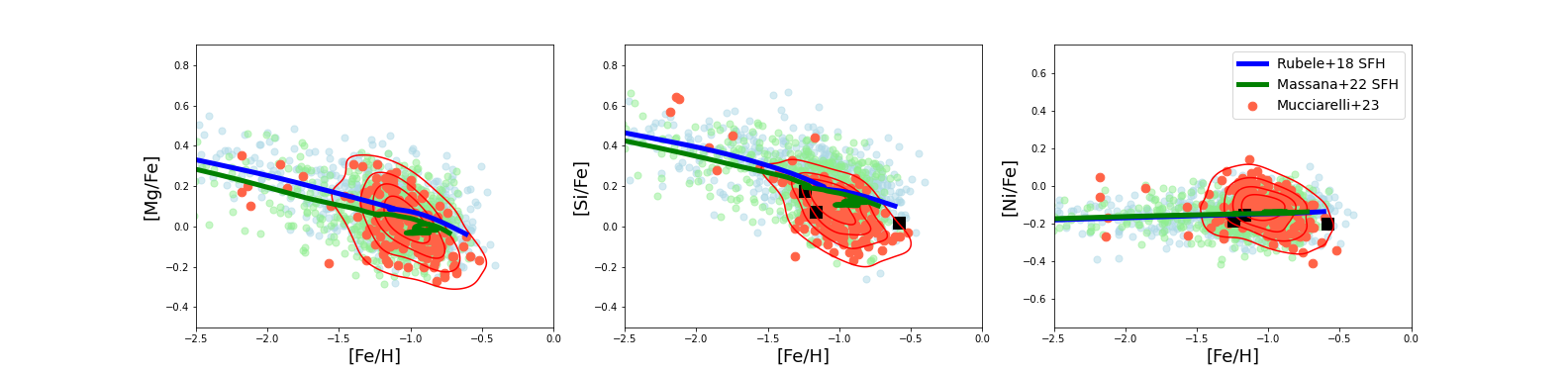}
    \caption{[X/Fe] vs. [Fe/H] for light elements in the SMC. Abundance patterns are shown for Mg (left panel), Si (central panel) and Ni (right panel) for reference models adopting \citet{Rubele18} SFH (blue lines and cyan regions) and \citet{Massana22} SFH (green lines light-green regions). 
    Solid lines are genuine chemical evolution tracks for models with a given SFH, whereas shaded regions are associated model predictions account for observational uncertainties ('synthetic model'). 
    Data are from \citet[][SMC field stars, orange dots]{Mucciarelli23} and \citet[][SMC GCs, black squares]{Mucciarelli23_GCs}. Orange contour lines represent density lines of the observed stellar distributions in SMC field stars.}
    \label{fig:XFe_FeH}
\end{figure*}

The comparison between the predictions of the models adopting \citetalias{Rubele18} SFH and \citetalias{Massana22} SFH and the observed abundance trends for Mg, Si, and Ni are shown in Fig. \ref{fig:XFe_FeH} (left, central and right panels, respectively). 
Both models show a good agreement with the bulk of the observed abundances: the predictions of our synthetic models (see Section \ref{ss:PhysQuant_MDF_ageFeH}) well cover the regions representing most of the data (red contours in Fig. \ref{fig:XFe_FeH}). 
Indeed, it is worth noting that the predicted [X/Fe] vs. [Fe/H]  patterns for the models adopting \citetalias{Rubele18} and \citetalias{Massana22} SFHs are basically indistinguishable when considering abundance uncertainties, with (rather small) differences arising only when looking at the homogeneous model tracks (see discussion in \citealt{Romano15}).
As a further check for the models, the predicted abundances are also compared with SMC GCs high-resolution observations by \citetalias{Mucciarelli23_GCs}: 
[Si/Fe] and [Ni/Fe] abundances (we do not consider Mg due to the Mg-Al anticorrelation in GCs, \citealt{Gratton12}) align well with the model outcome, confirming what already been seen for SMC field stars.

Therefore, despite small residual differences between model predictions and observations, the average age-[Fe/H], MDF, and [X/Fe] vs. [Fe/H] patterns are well reproduced, indicating that we have built a solid framework to probe the evolution of n-capture elements in the SMC (see next section). 
Indeed, the understanding of the nucleosynthetic origin of such elements is still limited (e.g. \citealt{Molero2023MNRAS.523.2974M,Molero25,Palla25}), making uncertainties about elemental nucleosynthesis much more relevant than the small discrepancies encountered in this section.


\section{n-capture elements in the SMC}
\label{s:ncapture_results}

Here, we focus on the evolution of n-capture elements Zr, Ba, La and Eu, obtained in the observations described in Section \ref{s:data}.
It is important to note that the four elements belong to different families, to which correspond different production mechanisms: Zr (Z=40) belongs to the first s-process peak, for which the predominant mechanism is the weak s-process in rotating massive stars \citep[e.g.][]{Pignatari10,Limongi18}; Ba (Z=56) and La (Z=57) are second peak s-process elements, where the main s-process mechanism in low-mass AGB stars is predominant \citep[e.g.][]{Cristallo15}; Eu (Z=63) is a pure r-process element (e.g. \citealt{Burris00,Sneden08}).

Therefore, the comparison performed in this Section represents the first attempt to model a representative number of n-capture elements with different origin simultaneously in the SMC, and in general in LG galaxies beyond the metallicity regime dominated by stochastic enrichment, namely at [Fe/H]$\lesssim -2$ (e.g. \citealt{Schoenrich19}). 
Indeed, previous theoretical works on LG dwarfs focus on Ba or Eu alone (e.g. \citealt{Hirai2015,Vincenzo15}) or Ba and Eu at most \citep[][for Fornax, Sculptors dSphs and Reticulum II UFD]{Molero21}. 
Furthermore, earlier works were based on older prescriptions concerning s- and r-process nucleosynthesis, which have received several revisions during recent years with relevant effects on well-known Galactic patterns (see, e.g. \citealt{Prantzos20,Molero2023MNRAS.523.2974M}).\\

\begin{figure*}
    \centering
    \includegraphics[width=1.0\textwidth]{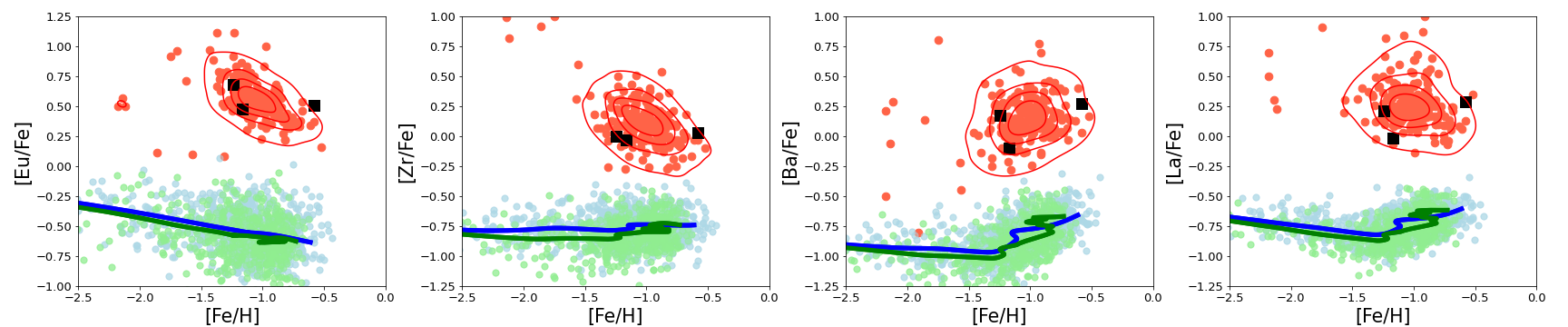}
    \caption{[X/Fe] vs. [Fe/H] for n-capture elements in the SMC. Abundance patterns are shown for Eu (leftmost panel), Zr (second leftmost) and Ba (second rightmost) and La (rightmost). 
    Solid lines are genuine chemical evolution tracks for the reference models adopting \citet{Rubele18} SFH (blue) and \citet{Massana22} SFH (green), whereas shaded cyan and light-green regions are associated model predictions account for observational uncertainties ('synthetic models'). 
    Data are from \citet[][Eu]{Anoardo25} and \citet[][Zr, Ba, La]{Mucciarelli23} for SMC field stars (orange dots) and \citet[][]{Mucciarelli23_GCs} for SMC GCs (black dots). Orange contour lines represent density lines of the observed stellar distributions in SMC field stars.}
    \label{fig:ncFe_std}
\end{figure*}

In Fig. \ref{fig:ncFe_std}, we show the model predictions for n-capture elements for both the models using \citetalias{Rubele18} and \citetalias{Massana22} SFHs, adopting the n-capture element setup as described in \citet{Molero2023MNRAS.523.2974M}. This setup (r-process sources, yields and delay times, s-process sources and yields, see also Section \ref{s:model}) allows one to reproduce the Galactic observables, from abundance patterns in r- and s-process elements to the measured MW event rates (see \citealt{Molero2023MNRAS.523.2974M,Palla25}).

Despite this, Fig. \ref{fig:ncFe_std} shows clearly that the adopted enrichment setup for n-capture elements fails to reproduce the observed patterns in the SMC. 
Indeed, all the elements under scrutiny are severely underestimated, even accounting for abundance uncertainties with our synthetic framework. 
In addition, models adopting different SFHs show minimal variations in terms of their abundance patterns for all these elements. For this reason, during the rest of the paper we plot only the model predictions obtained from \citetalias{Rubele18} SFH (but see Fig. \ref{fig:ncFe_moreMNS_compareSFH} in Appendix), as to facilitate the comparison between data and models. 
It is worth noting that a deficiency in Eu production in the LG by GCE models was noted in \citet{Palla25} for Sagittarius, Fornax and Sculptor dwarf galaxies. However, here we note that a similar deficiency is also present for other n-capture elements for which the s-process should be the main production mechanism \citep[e.g.][]{Prantzos20}.
Therefore, in the following we try to reconstruct the origin of this discrepancy by analysing several pathways to recover such a missing production.

\subsection{Increased r-process production at low-metallicity}
\label{ss:moreEu}

As a first step, we focus on Eu which is a pure r-process element, namely is (almost) only produced by means of the r-process \citep{Burris00,Sneden08}.
In this way, we isolate one production channel that however has a non-negligible contribution in the nucleosynthesis of Zr, Ba and La ($\simeq$19\%, 14\%, 24\% at solar, \citealt{Sneden08}).

As already mentioned, \citet{Palla25} noticed a lack in [Eu/Fe] ($\gtrsim$ 0.5 dex) for GCE models of several LG dwarfs, suggesting that 3/4 of the needed Eu production in dwarf galaxies is missing, despite the adopted prescriptions reproduce the observed MW patterns. To reconcile the discrepancy between the Galactic and extragalactic observations, they suggested a marked increase in Eu production by MNS events at low metallicity. 
In particular, they adopt a variable fraction of stars that originate MNS events ($\alpha_{MNS}$), in this way:
\begin{equation}
    \alpha_{MNS} = \left\{
    \begin{array}{ll}
    \alpha_{MNS,0} \times 40 \hspace{0.75cm} {\rm if} \, \, {\rm Z}\leq {\rm Z_{thresh}}\\[0.1cm]
    \alpha_{MNS,0} \hspace{1.5cm} {\rm if} \, \, {\rm Z} > {\rm Z_{thresh}}\\
    \end{array}, \right.
    \label{eq:alpha_incr}
\end{equation}
where $\alpha_{MNS,0}=2\cdot 10^{-3}$ and ${\rm Z_{thresh}}= 2\cdot10^{-3}\simeq 0.1 \ {\rm Z_\odot}$.\\

\begin{figure*}
    \centering
    \includegraphics[width=1.0\linewidth]{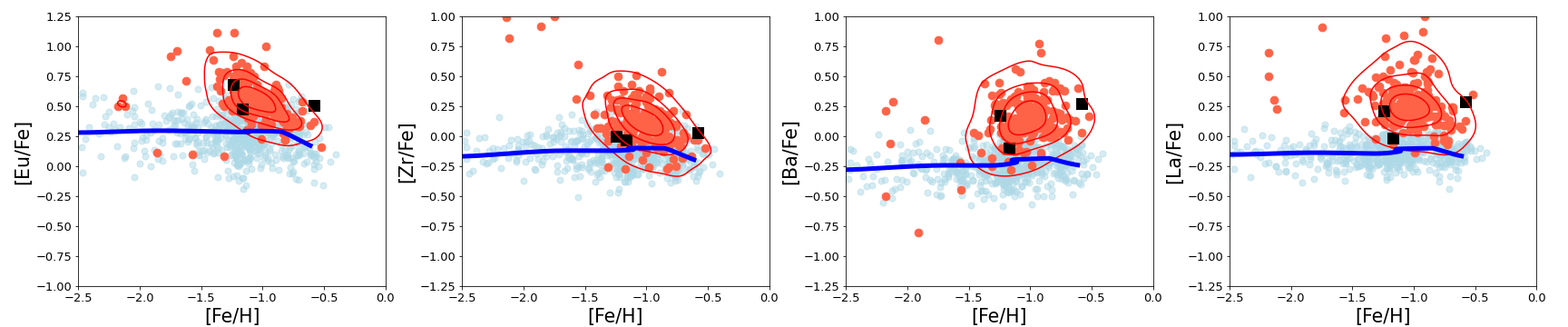}
    \caption{[X/Fe] vs. [Fe/H] for n-capture elements for the model adopting \citet{Rubele18} SFH with increased MNS r-process production at low-metallicity (see Eq. \eqref{eq:alpha_incr}). Blue solid lines and cyan shaded regions represent genuine chemical tracks and model predictions accounting for observational uncertainties ('synthetic model'), respectively. Data are as in Fig. \ref{fig:ncFe_std}.}
    \label{fig:ncFe_moreMNS}
\end{figure*}

Therefore, in Fig. \ref{fig:ncFe_moreMNS} we apply the same formulation as in Eq. \eqref{eq:alpha_incr} to our chemical evolution model for the SMC. 
As we note in the Figure left panel, the boost in Eu production by MNS produces a significant rise in the [Eu/Fe] abundance, with a plateau extending up to metallicities slightly above [Fe/H]=-1 dex. 
The extended plateau is the effect of the large Eu production rate by MNS originating up to the given metallicity threshold. 
After the metallicity threshold for enhanced Eu production is passed and the enhanced production progressively fades, the [Eu/Fe] ratio declines with metallicity due to "unbalanced" production of Fe (mainly by Type Ia SNe) relative to Eu.
In any case, the new predicted trend for Eu remains in part below the bulk of SMC stars, especially for metallicities below [Fe/H]$\sim -1$ dex.

A significant reduction of the offset observed in Fig. \ref{fig:ncFe_std} is also seen for the three s-process elements in the three rightmost panels of Fig. \ref{fig:ncFe_moreMNS}. 
Despite being dubbed as s-process elements, Zr, Ba and La receive a fundamental contribution from r-process at low-metallicities, where s-process production is disfavoured by the low-metallicity environment (due its secondary nature\footnote{element production stems from metal seeds already present in the star at birth, namely the element production increases with metallicity.}) and/or the longer lifetimes of the stellar production sources (low-mass stars).
However, especially for Ba and La where the s-process production is dominated by low-mass stars (main s-process)
, we notice a clear [X/Fe] deficit in the predictions: the model does not show the expected increase in the [s-process/Fe] ratio driven by low-mass stars at high-metallicity \citep[][and references therein]{Molero2023MNRAS.523.2974M}, exhibiting instead a flat trend that prevents to cover most of the SMC field stars and GCs data. 
It is worth noting that for all the n-capture abundance patterns in Fig. \ref{fig:ncFe_moreMNS}, analogue results are found for the \citetalias{Massana22} SFH, as shown by Fig. \ref{fig:ncFe_moreMNS_compareSFH} in Appendix. This further confirms the small differences in terms of [X/Fe] vs. [Fe/H] patterns by the two SFHs, even in the presence of nucleosynthesis prescriptions dependent on galactic SF and chemical enrichment.\\

In summary, the increase in MNS r-process production at low metallicities  as proposed in \citet{Palla25} clearly helps to reconcile the predictions with observations of n-capture elements. 
However, we still fall short in explaining several of the features described by the observed abundance patterns in the galaxy.

\subsection{Impact of the IMF}
\label{ss:IMF}

Until now, we have adopted the IMF by \citet{Kroupa01} to be consistent with the adopted SFHs, derived by assuming the same IMF. 
However, it is worth exploring whether variations in the IMF can significantly influence the evolution of n-capture elements in the SMC. Indeed, the IMF regulates the relative fractions of low- and high-mass stars within stellar populations, which  has a critical impact on the metal production and abundance ratios in galaxies.
In addition, several indications \citep[e.g.][]{Lee09,Watts18} point towards steeper (or top-light) IMFs in the high-mass domain in dwarf galaxies, namely reducing the relative fraction of massive stars. These can be framed in the context of the integrated galactic IMF (IGIMF) theory, where galaxies with lower rates of star formation are prone to underproduce massive stars (e.g. \citealt{Jerabkova18,Yan20}).

Therefore, in this Section we test the effect of different IMFs from the canonical one (\citealt{Kroupa01}, high-mass end slope $x=-1.3$), by adopting the \citet{Kroupa93} IMF formulation ($x=-1.7$) or a modification of the latter with an even steeper slope (hereafter dubbed as top-light, $x = -2$). 
It is worth remarking that our test does not have the intention to dig deeper into the intricate debate about the IMF (see \citealt{Jerabkova25}) and we know that different assumptions in the IMF can produce differences in the extracted SFHs. However, Figs. \ref{fig:XFe_FeH} and \ref{fig:ncFe_std} demonstrate relatively small differences in the abundance patterns obtained by means of different SFHs for the galaxy, with other ingredients being more relevant in shaping [X/Fe] vs. [Fe/H] diagrams.
In any case, since the change in the IMF has consequences not only on the production of n-capture elements but also on the global metal evolution, all the models adopting an IMF different from the canonical one are re-calibrated in the SFE $\nu$ to reproduce the main SMC observables as in Section \ref{s:results} (see Appendix \ref{a:IMF_calibration}). In this way, we avoid spurious results arising from chemical evolution histories inconsistent with observations.\\

\begin{figure*}
    \centering
    \includegraphics[width=1.0\linewidth]{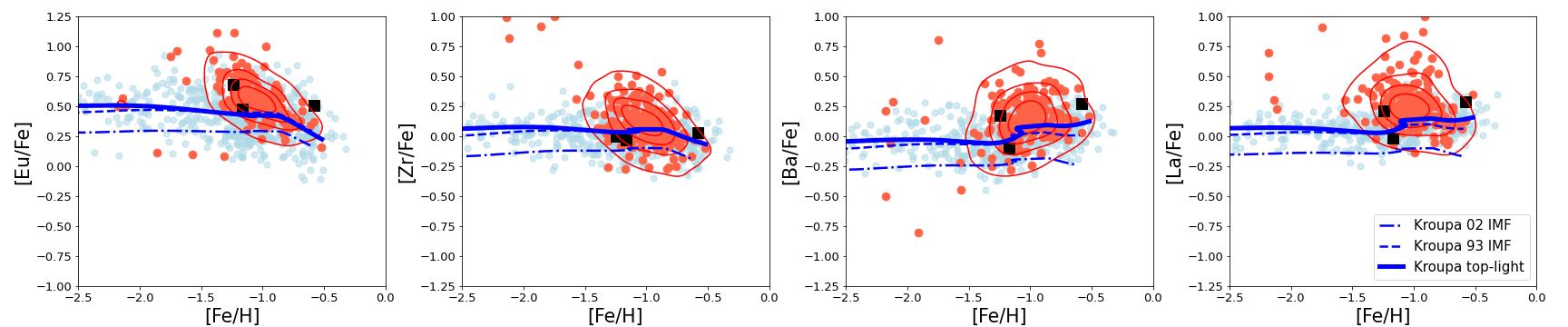}
    \caption{[X/Fe] vs. [Fe/H] for n-capture elements for the model with \citet{Rubele18} SFH and increased MNS r-process production at low-metallicity (see Eq. \eqref{eq:alpha_incr}), adopting different IMFs. Blue lines represent models adopting the reference \citet{Kroupa01} IMF (thin dash-dotted), \citet{Kroupa93} IMF (thin dashed) and a modified \citet{Kroupa93} IMF with steeper high-mass end slope (top-light IMF, thick solid). Cyan shaded regions are predictions for the model with top-light IMF accounting for observational uncertainties ('synthetic model'). Data are as in Fig. \ref{fig:ncFe_std}.}
    \label{fig:ncFe_moreMNS_IMF}
\end{figure*}

The [X/Fe] vs. [Fe/H] patterns for Eu, Zr, Ba and La for the models with enhanced MNS r-process production adopting the \citet{Kroupa93} IMF and the top-light IMF are shown in Fig. \ref{fig:ncFe_moreMNS_IMF}, together with the one assuming the canonical \citet{Kroupa01} IMF. For the model with the top-light IMF, we also show the prediction from the synthetic framework (see Section \ref{s:results}).

The Figure shows that the proposed change in the IMF has a beneficial effect in the reproduction of the abundance patterns. 
For [Eu/Fe] (Fig. \ref{fig:ncFe_moreMNS_IMF} leftmost panel), the predicted behaviour for the model with top-light IMF remains similar to the one shown in Fig. \ref{fig:ncFe_moreMNS}. However, the plateau at lower metallicites is shifted to higher values ($\sim 0.5$ dex, difference $\sim 0.25$ dex from the model with the canonical IMF). 
This happens because the models adopting top-lighter IMFs need for larger SFE to reproduce the SMC observables, increasing the contribution to r-process enrichment by MRD-SNe. In turn, when abundance uncertainties are considered, this allows us to reproduce most of the observed field stars and GCs in the SMC.

For [Zr/Fe] (Fig. \ref{fig:ncFe_moreMNS_IMF} second leftmost panel), the predictions from the models show an extended plateau similar to Eu, but at roughly solar [Zr/Fe] values for the models adopting the steeper IMFs. Then the models show a small rise between [Fe/H]$\sim$-1.25/-1 dex, due to increased s-process production mostly by (rotating) massive stars caused by the peak in SF at intermediate ages (see Fig. \ref{fig:SFHs}), followed by a gentle decline at the highest metallicities as a consequence of Type Ia SN enrichment. Also in this case, the adoption of a steeper IMF greatly helps ro reproduce the general trend captured by the data. Again, this happens for the rise in SFE for models with steeper IMF slopes, which in this case increases not only the r-process with MRD-SNe but also the weak s-process channel through massive stars.

However, it is for Ba (Fig. \ref{fig:ncFe_moreMNS_IMF} second rightmost panel) that the adoption of an IMF favouring/disfavouring the presence of low-mass/massive stars is the most beneficial. In fact, the top-light IMF not only allows for an upward shift in the low-metallicity plateau, but also for a rise in [Ba/Fe] from metallicties [Fe/H]$\gtrsim-1.25$ dex. The latter is the direct consequence of the larger weight of low-mass stars in chemical enrichment, as Ba (and La as well) is preferentially produced through the main s-process in low-mass ($\sim$1-3 M$_\odot$) stars.  
This increase in [Ba/Fe] abundance with metallicity, despite being not very large ($\sim$ 0.25 dex), is crucial to recover a good agreement with the bulk of field star and GC data, which otherwise would remain underestimated by models.
The rise with metallicity offers an important aid in the data-model agreement also for [La/Fe] (Fig. \ref{fig:ncFe_moreMNS_IMF} rightmost panel). However, despite the substantial improvement in reproducing the observed La, part of the data remains above the predictions, even accounting for observational errors.  \\

\begin{figure*}
    \centering
    \includegraphics[width=0.925\linewidth]{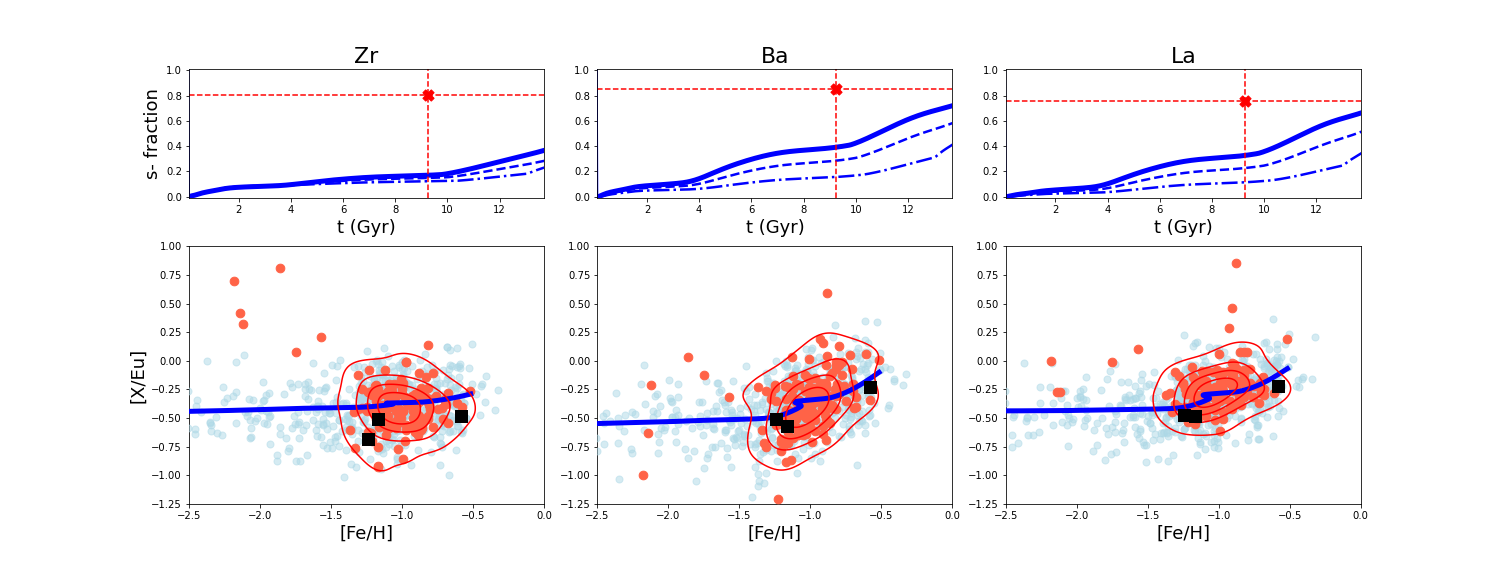}
    \caption{Mass fraction produced by the s-process channel of production (top panels) and [X/Eu] vs. [Fe/H] (bottom panels) for Zr (left panels), Ba (central panels) and La (right panels) for models with different IMFs.
    In top panels, we show results for models adopting the reference \citet{Kroupa01} IMF (thin dash-dotted lines), \citet{Kroupa93} IMF (thin dashed) and a modified a top-light IMF (thick solid). 
    In bottom panels, blue solid lines an cyan shaded regions are genuine chemical tracks and predictions accounting for observational uncertainties ('synthetic model') for the model with top-light IMF. 
    Data are Solar s-process mass fraction \citep[][top panels]{Sneden08}  and as in Fig. \ref{fig:ncFe_std} (bottom panels).}
    \label{fig:ncEu_sfrac}
\end{figure*}

To probe in more detail the evolution of heavy elements and in particular the relative contribution between the r-process and s-process channels, in Fig. \ref{fig:ncEu_sfrac} we show the predicted mass fraction evolution of s-process component (top panels) and [X/Eu] abundance ratios as a function of metallicity compared with the samples described in Section \ref{s:data} (bottom panels).
Fig. \ref{fig:ncEu_sfrac} bottom panels show a good agreement between observation and predictions. The model adopting a top-light IMF nicely reproduces the trends observed in [Zr,Ba,La/Eu], with data superimposed to predictions in our synthetic GCE framework.

In particular, Zr shows a flat/slowly increasing trend relative to Eu, denoting a relevant r-process contribution throughout galactic evolution. Indeed, the s-process fraction evolution for Zr (top left panel in Fig. \ref{fig:ncEu_sfrac}) denotes a subdominant production through the s-process channel, with a mass fraction of the order of 0.4 at the present-day.
In contrast, we observe larger s-process contribution for Ba and La (two rightmost panels on top) with present-day values of $\sim 0.7-0.8$ for both elements, obtained by a progressive increase throughout Gyrs. 
The increase in s-fraction is also visible in the [Ba,La/Eu] vs. [Fe/H] patterns in Fig. \ref{fig:ncEu_sfrac} bottom panels, with a sharp rise in [X/Eu] starting from [Fe/H]$\sim$-1.25 dex up to the highest predicted metallicities. This is a consequence of delayed enrichment by low-mass stars, which contribution is the dominant channel for main s-process elements (as Ba and La). 

This is also demonstrated by the lower s-process fractions by models adopting an IMF favouring massive stars production (see dashed and dash-dotted lines in the upper panels). For Ba and La, the s-channel contribution at present-day is reduced by more than 30\% by adopting a \citet{Kroupa01} IMF, almost halving the role of the s-process in the synthesis of elements. 
A reduction in the s-process fraction is also seen for Zr, but here of a 10$\%$ only. Moreover, the decrease for Zr is mostly driven by the lower contribution from massive stars, as due to the lower SFE in the models adopting top-heavier IMFs (see earlier in the Section).

\subsection{Different IRV distribution for massive stars}
\label{ss:different_IRV}

Other possible explanations for the observed abundance patterns in n-capture elements can also be advocated.
In particular for s-process, the initial rotation velocity (IRV) distribution in massive star populations is an active object of study \citep[e.g.][]{Rizzuti19,Prantzos20,Molero25}.
For our reference yield set for massive star s-process production (\citealt{Limongi18} set R150) we adopted a fixed initial rotational velocity of 150 km s$^{-1}$. Even though such a prescription is observationally motivated, as it allows reproducing Galactic observations for a large number of n-capture elements (see \citealt{Molero2023MNRAS.523.2974M}), the adoption of sets with different rotational velocities (for \citealt{Limongi18}, $v_{rot}=0, 300$ km s$^{-1}$) or a distribution of velocities is worth to be considered.

Therefore, beside the reference yield set R150, in the following additional sets are considered.
We test individual sets (\citealt{Limongi18} set R300, $v_{rot}= 300$ km s$^{-1}$), or distributions of rotational velocities varying with metallicity. For the latter, we adopt the distributions as in \citet[][hereafter DIS2]{Molero24} and \citet[][hereafter DIS3]{Romano19}, both assuming that the probability that a star rotates at a certain speed is a function of the metallicity, Z, with faster velocities being more likely at lower Z (see Tab. \ref{tab:IRV} for details). 
This is supported by the expectation that massive stars rotate faster at lower metallicities, where they are more compact (see \citealt{Klencki20}). This view is supported both theoretically and observationally (e.g. \citealt{Frischknecht16,Martayan07a,Martayan07b,Hunter08}, see \citealt{Molero24} for a discussion). 
It is worth mentioning that other IRVs to those shown in Tab. \ref{tab:IRV} can be considered. However, exploring in detail the IRV of massive stars in the SMC goes beyond the scope of this paper, with other literature distributions either showing more extreme [X/Fe] patterns (at odds with observations) or very similar results to the setups displayed in Tab. \ref{tab:IRV}.\\

\begin{table}[]
    \centering
    \caption{IRV distributions for massive stars (\citealt{Limongi18} models, set R) tested in this study in addition to the reference set (R150, see Section \ref{ss:nucleosynthesis}).}
    \begin{tabular}{c| c}
        \hline
        \multirow{2}{*}{Model}  &  IRV distribution \Tstrut \\
        &  (\% at [Fe/H]=-3, -2, -1, 0) \Bstrut\\
        \hline
        R300 &  300 km s$^{-1}$: 100, 100, 100, 100 \Tstrut\\[0.2cm]
        \multirow{2}{*}{DIS2} &  0 km s$^{-1}$: 10, 20, 30, 45\\
        \multirow{2}{*}{\citep{Molero24}}& 150 km s$^{-1}$: 10, 30, 50, 50\\
         & 300 km s$^{-1}$: 80, 50, 20, 5\\[0.2cm]
        DIS3  & 0  km s$^{-1}$: 0, 0, 100, 100\\
        \citep{Romano19}& 300 km s$^{-1}$: 100, 100, 0, 0 \Bstrut\\
        \hline
    \end{tabular}
    \label{tab:IRV}
\end{table}

\begin{figure*}
    \centering
    \includegraphics[width=0.925\linewidth]{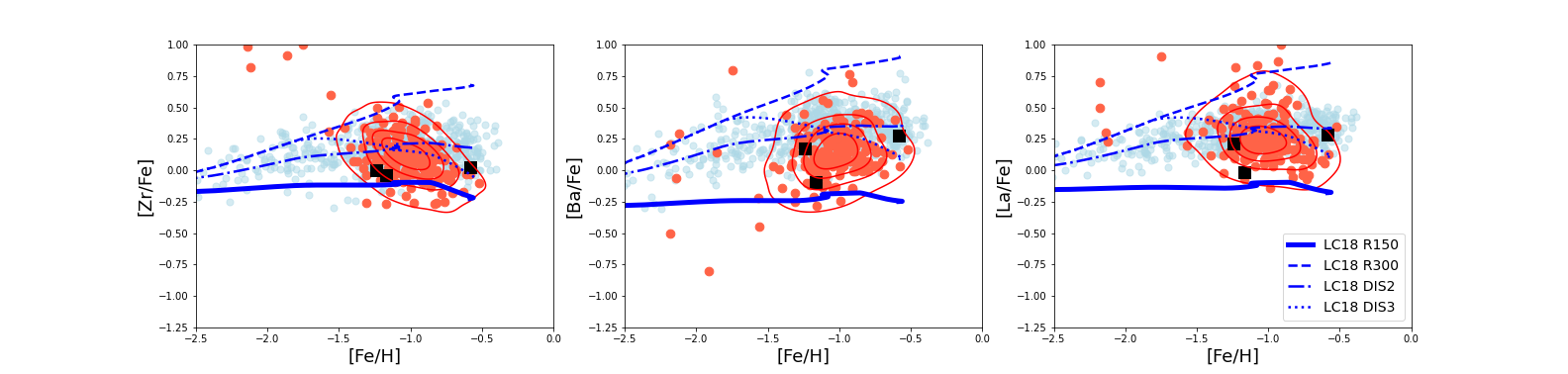}
    \caption{[X/Fe] vs. [Fe/H] for Zr, Ba and La for the model with \citet{Rubele18} SFH and increased MNS r-process production at low-metallicity (see Eq. (4)), adopting different IRV distributions (see Tab. \ref{tab:IRV}). 
    Blue lines represent models adopting the reference IRV distribution (R150, thick solid lines), R300 distribution (thin dashed), DIS2 distribution (thin dash-dotted) and DIS3 distribution (thin dotted). 
    Cyan shaded regions are predictions for the model with the DIS2 distribution accounting for observational uncertainties ('synthetic model’). 
    Data are as in Fig. \ref{fig:ncFe_std}.}
    \label{fig:different_IRVs}
\end{figure*}

The predicted [Zr,Ba,La/Fe] vs. [Fe/H] chemical tracks for models adopting a \citet{Kroupa01} IMF and the massive stars IRV distributions in Tab. \ref{tab:IRV} are shown in Fig. \ref{fig:different_IRVs}: the models with R300, DIS2 and DIS3 distributions show enhanced [X/Fe] relative to the reference yields adopted in this paper. 
Going more into detail, the model with all stars rotating at $v_{rot}=300 \ {\rm km \ s^{-1}}$ (R300, dashed lines)  overestimates the predicted trends relative to the observations, indicating an excess of s-process production by high-speed rotators when assumed as the sole contributors to the weak s-process channel.
Indeed, when assuming that only a fraction (decreasing with metallicity) of massive stars are high-velocity stellar rotators (DIS2, dash-dotted lines) the [X/Fe] enrichment level generally decreases, with predictions more in agreement with observations especially for Zr and La. 
A fairly good agreement between models and observations can also be seen for the model adopting the DIS3 IRV distribution (dotted lines), where a decreasing trend in the [s-process/Fe] ratios is observed starting from [Fe/H]$\simeq-2$ dex. This is due to the progressive cease of contribution from $v_{rot}=300 \ {\rm km \ s^{-1}}$ rotators for metallicities beyond [Fe/H]$=-2$ dex. In turn, this allows one to reproduce well the observed decreasing [Zr/Fe] trend, whereas it is at odds with the observed increase in Ba with metallicity.

\begin{figure*}
    \centering
    \includegraphics[width=0.925\linewidth]{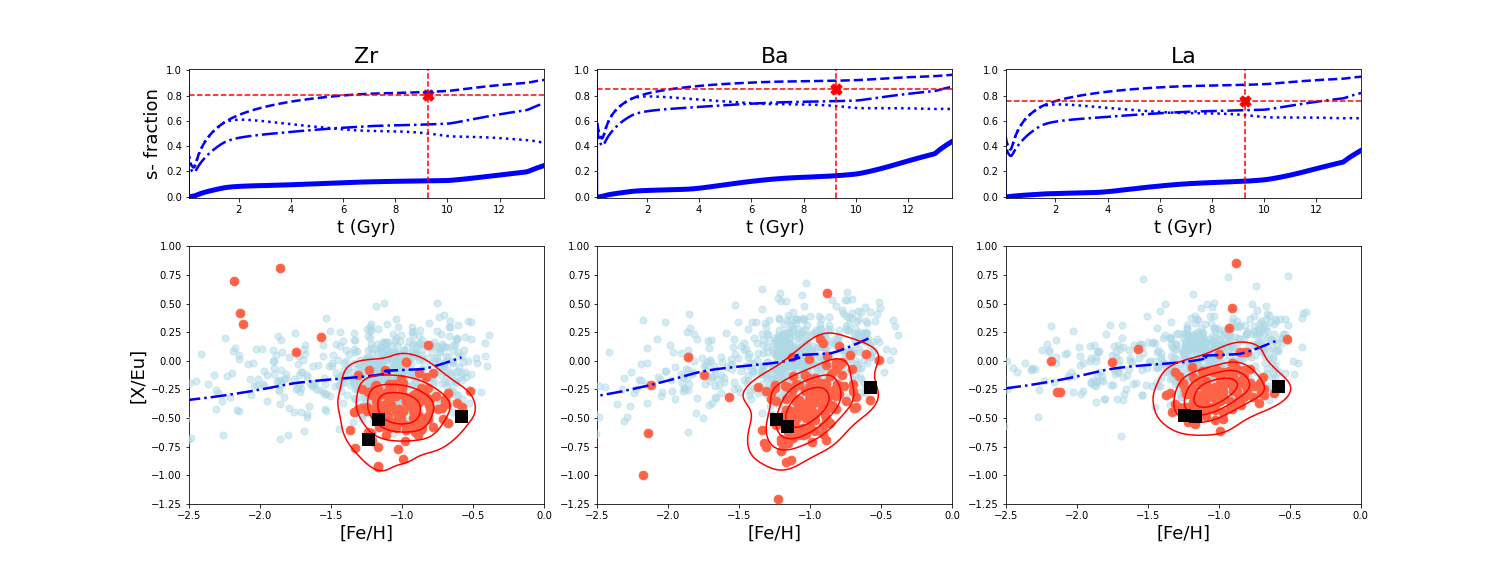}
    \caption{Mass fraction produced by the s-process channel of production (top panels) and [X/Eu] vs. [Fe/H] (bottom panels) for Zr (left panels), Ba (central panels) and La (right panels) for models with different IRV distributions.
    In top panels, we show results for models adopting the reference IRV distribution (R150, thick solid lines), R300 distribution (thin dashed), DIS2 distribution (thin dash-dotted) and DIS3 distribution (thin dotted). 
    In bottom panels, blue dash-dotted lines and cyan shaded regions are genuine chemical tracks and predictions accounting for observational uncertainties ('synthetic model') for the model with the DIS2 distribution. Data are as in Fig. \ref{fig:ncEu_sfrac}.}
    \label{fig:different_IRVs_sfrac}
\end{figure*}

As highlighted in Section \ref{ss:IMF}, a more detailed look can be achieved by looking at [X/Eu], a proxy to indicate the fractional contribution from the s- and r- process throughout galactic evolution. For this reason, in Fig. \ref{fig:different_IRVs_sfrac} we show the [X/Eu] abundance ratios as function of the metallicity (bottom panels) and the mass fraction evolution of s-process (top panels) for Zr, Ba and La.
All models display very high s-process mass fractions with values $\gtrsim 0.6$ already at 2 Gyr of evolution, especially for Ba and La. This is also reflected in the predicted [X/Eu] patterns (shown in Fig. \ref{fig:different_IRVs_sfrac} for the DIS2 distribution) which lie above the abundances observed in the SMC stars. Indeed, even by post-processing the output of the model with abundance uncertainties, the predictions are only able to reproduce the upper envelope of the data.  
Similar or even worse comparisons in abundance patterns are observed for the other distributions tested, reaching an overestimation in [X/Eu] of the order of 0.5-0.75 dex relative to the observations in the case of the R300 model, which shows the largest s-process fraction ($\gtrsim 0.8$ at 2 Gyr of evolution).\\

In summary, the contribution of high speed rotators ($v_{rot}=300 \ {\rm km \ s^{-1}}$) should be marginal in order to reproduce the observed trends in the SMC, at least with the yield sets by \citet{Limongi18}. 
This also happens using the other IMFs tested in the previous Section (\citealt{Kroupa93}, top-light). In fact, despite not favouring the presence of massive stars in a stellar population as for the reference IMF, the larger SFE adopted for steeper IMF slopes (see \ref{ss:IMF}) has the effect in maintaining constant, where not increasing, the contribution from rotating massive stars to the s-process enrichment (see Fig. \ref{fig:IMF_IRV} in Appendix).

\subsection{i-process from low-mass stars}
\label{ss:iproc}

In addition to r- and s-processes, other secondary neutron capture processes have been identified. The so-called intermediate-process \citep[i-process,][]{Cowan77}, taking place at intermediate neutron irradiation between r- and s-processes, has been claimed several times to explain stellar overabundances not compatible with the s- or the r-processes alone (e.g. \citealt{Roederer16,Mashonkina23,Hansen23}). 
Theoretically, i-process can develop if protons are mixed in a convective helium-burning zone (proton ingestion event, PIE). This could happen in different astrophysical sites (see e.g. \citealt{Choplin21}) and in particular in low-metallicity, low-mass AGB stars (e.g. \citealt{Iwa04,Cristallo09,Choplin21}).
As in the SMC we are dealing with a low-metallicity environment, it is worth exploring such a channel: in fact, the chemical imprint by this nucleosynthetic process can be more easily detected in such metal-poor, slower star forming systems.

To this aim, we implement the AGB yields including PIEs by \citet{Choplin22,Choplin24} for stars spanning masses between $1 \leq m/$M$_\odot \leq 3$ and metallicities between $-3<$[Fe/H]/dex$<-0.5$, by replacing the reference FRUITY yields in the above mentioned ranges.
To obtain the densest gridding in terms of masses and metallicities, we adopt models without overshooting for [Fe/H]$\leq-2.5$ dex, while for larger metallicities models with overshooting parameter at the top of AGB thermal pulse $f_{top}=0.1$ 
(see \citealt{Choplin22}).\\

\begin{figure*}
    \centering
    \includegraphics[width=1.0\linewidth]{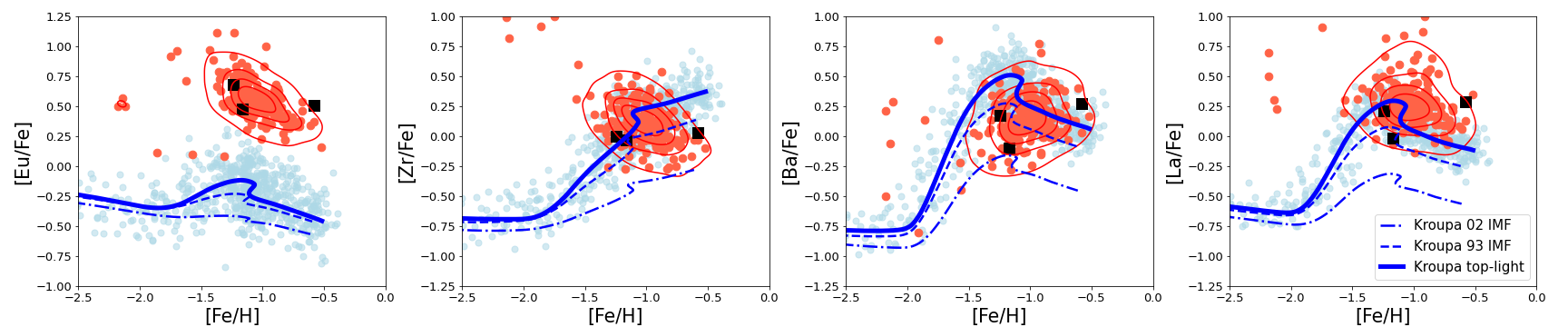}
    \caption{[X/Fe] vs. [Fe/H] for n-capture elements for the model with \citet{Rubele18} SFH and low-mass AGB yields by \citet{Choplin22,Choplin24}, adopting different IMFs. Blue lines represent models adopting the reference \citet{Kroupa01} IMF (thin dash-dotted), \citet{Kroupa93} IMF (thin dashed) and a modified \citet{Kroupa93} IMF with steeper high-mass end slope (top-light IMF, thick solid). Cyan shaded regions are predictions for the model with top-light IMF accounting for observational uncertainties ('synthetic model'). Data are as in Fig. \ref{fig:ncFe_std}.}
    \label{fig:ncFe_iproc_IMF}
\end{figure*}

In Fig. \ref{fig:ncFe_iproc_IMF}, we present the n-capture abundance patterns for chemical evolution models without increased MNS r-process production and the yields for low-mass stars as in \citet{Choplin22,Choplin24} for different IMFs. The choice of not showing models with an increased r-process production rate at low-metallicity can be view in the light of i-process being proposed as a possible prolific n-capture element source at low-metallicity (e.g. \citealt{Hompel16,Choplin24b} and references therein).

Looking at the figure, we first notice that the top-lighter the IMF, the larger the predicted [X/Fe] ratios across different metallicities, as due to the more favourable conditions imposed for the formation of low-mass stars. 
However, Fig. \ref{fig:ncFe_iproc_IMF} left panel shows that despite some production of Eu by means of low-mass AGBs through the i-process at metallicities between -2$\lesssim{\rm [Fe/H]/dex}\lesssim$-1, this is not enough to fulfil the budget as required by observations. This finding reinforces the claim of a strong increase in the r-process production rate at low-metallicity (see \ref{ss:moreEu}). Regarding the predicted bump in [Eu/Fe], we note that it is also present for Ba and La. 
Indeed, yield tables for Eu, Ba and La show very large [X/Fe] values for metallicities between [Fe/H]$\simeq$-2.5 dex and [Fe/H]$\simeq$-1.5 dex (see Fig. 8 in \citealt{Choplin24b}), whereas significantly lower yields (even subsolar in [X/Fe]) are found at larger metallicities. The larger metallicities at which the bump is obtained in the SMC chemical evolution are due to the significant delay in enrichment by stars with low masses ($\gtrsim$1 Gyr for  masses $\lesssim$2.5 M$_\odot$). 
For Zr instead, a progressive increase in [X/Fe] with metallicity is observed. This happens because the low-mass star yields remain large up to metallicties as [Fe/H]$\simeq$-1 dex for the set adopted in this Section.
Such metallicities are reached only at late times in the SMC evolution (see Fig. \ref{fig:AgeFeH_MDF_Rubele}), preventing bump signatures due to pollution time delay. The different behaviour for Zr 
can be explained in light of their different position in the periodic table: Zr belongs to the so-called n-capture 1st peak, at variance with the other elements (belonging to 2nd peak). The former meets favourable conditions for production even at smaller neutron-to-seed ratios, allowing a broader metallicity range at which i-process nucleosynthesis is effective (for similar neutron densities, the higher the metallicity the smaller the neutron-to-seed, see \citealt{Choplin24b}).

In general, however, the model predictions including i-process nucleosynthesis by low-mass stars struggle to reproduce the observed trends, with model tracks showing increasing trends with metallicities where stars point towards decreasing trends (e.g. Zr) and vice versa (e.g. Ba).
Moreover, it is worth mentioning that the agreement does not improve significantly when imposing an additional r-process production at low-metallicity (see Fig. \ref{fig:ncFe_iproc_MNS} in Appendix). \\

Therefore, the analysis performed in this Section rejects the i-process nucleosynthesis in low-mass AGB stars as a major contributor for n-capture elements in the SMC.
Nonetheless, we cannot completely exclude this process as a contributor to the chemical enrichment. In fact, it is worth recalling the very large uncertainties in the modelling of this process, with poor constraints in several parameters crucially affecting its efficiency (e.g. overshooting, see \citealt{Choplin24}).

\section{Discussion and conclusions}
\label{s:discussion_conclusion}

In this study, we investigate the chemical enrichment of the SMC, the second largest satellite of the Milky Way (MW). 
To this aim, we develop chemical evolution models built upon inferred SFHs obtained by means of  colour-magnitude diagram fitting  (from \citealt{Rubele18,Massana22}), following in detail the chemical feedback provided by a wide variety of nucleosynthetic sources.

Within our model framework, we devote particular attention to the evolution of neutron-capture (n-capture) elements, which theoretical understanding has become a major topic in stellar physics and MW chemical evolution studies (see, e.g. \citealt{Cote19,Koba20all,Prantzos20,Molero2023MNRAS.523.2974M,Arconesreview23}), but which received fewer attention on the side of dwarf galaxies (see \citealt{Palla25}). In particular, we investigate the impact of several factors on the abundance patterns of these heavy elements, such as their production by means of the rapid n-capture (r-) process, the impact of the IMF and different IRV distributions for massive stars.
To gain insights on such properties, our models are compared with recent observational data for the SMC, 
including measurements for n-capture elements produced by different channels (\citealt{Mucciarelli23,Anoardo25}).\\

In summary, our main conclusions are:
\begin{itemize}
    \item despite some small variations in the predicted age-metallicity relation and the stellar metallicity distribution function, the two SFHs tested in this work \citep{Rubele18,Massana22} do not cause major differences in the [X/Fe] vs. [Fe/H] abundance patterns for the SMC.
    The small variations ($\lesssim 0.1$ dex) hold for elements from different nucleosynthetic sources and are well below typical modelling uncertainties (e.g. yields, delay time distribution functions) for the evolution of n-capture elements, which are the main focus of this study;

    \item models with n-capture enrichment prescriptions analogous to those adopted to reproduce the observed patterns in the MW disk \citep{Molero2023MNRAS.523.2974M} fail to reproduce the trends observed in the SMC. In particular, models significantly underestimate the production of these elements by showing a [X/Fe] deficiency  $\gtrsim 0.5$ dex. 
    This result aligns with the findings by \citet{Palla25}, showing a systematic underestimation of [Eu/Fe] in models of other Local Group dwarf galaxies relative to the observations. In this study, our findings extend to other n-capture elements that in the MW are primarily synthesised by means of the slow n-capture (s-) process (Zr, Ba, La). In turn, this highlights the primary importance of the r-process production channel for these elements in the SMC, especially at low metallicities;

    \item by assuming a large boost 
    in r-process production by delayed sources (i.e. compact binary mergers) at $Z< 0.1 $ Z$_\odot$ as suggested by \citet{Palla25}, the model predictions tend to reconcile with the observed trends for n-capture elements. This result confirms the need for additional r-process production at low metallicity to explain the abundance patterns in dwarf galaxies. 
    However, for the s-process elements Ba and La in particular, model tracks do not show the observed rise in the [X/Fe] in the metal rich end. 
    As these elements should be heavily synthesised in low-mass stars (main s-process, see e.g. \citealt{Karakas10,Cristallo15}), this points towards a stronger contribution by these stars to the chemical enrichment relative to our standard predictions;

    \item in light of the findings above, we test the effects of different IMFs. The adoption of a top-lighter (i.e. disfavouring massive stars) IMF relative to the reference one in this work \citep{Kroupa01} is highly beneficial in reproducing the trends for n-capture elements as observed in the galaxy.
    In particular, a \citet{Kroupa93} IMF with a steeper high-mass slope ($x=-2$, see Section \ref{ss:IMF}) allows [Ba/Fe] and [La/Fe] to rise with metallicity for [Fe/H]$\gtrsim -1.5$ dex, in better agreement with observations. 
    In addition, this top-light IMF generally produces a larger [n-capture/Fe] plateau at low metallicities, improving the agreement with observations especially for Zr and Eu. As a consequence, predictions for a top-light IMF also well reproduce the observed [s-process/Eu] trends, which can be used as proxies of the s-/r-process contribution fraction throughout the galaxy chemical evolution;

    \item in addition to our reference yield set for n-capture in massive stars \citep[R150]{Limongi18}, we test several IRV distributions from the literature (see Table \ref{tab:IRV}).
    We find that distributions favouring high-speed rotators ($v_{rot}= 300$ km s$^{-1}$) worsen the agreement with s-process element trends, especially Ba. Indeed, stellar yields for these objects show prolific s-process synthesis, leading to very large s-process budget in the early stages of galactic evolution (up to 80\% already at 2 Gyr) at variance with our reference model predictions.  
    This is also noted in the overabundance in the [X/Eu] ratios by models including high-speed rotators, indicating an excess in the s-process synthesis;

    \item the adoption of low-mass AGB yields that include i-process nucleosynthesis \citep{Choplin22,Choplin24} does not improve significantly the reproduction of r-/s-process element patterns. 
    While not producing a considerable rise in the predicted [Eu/Fe] to match the observations, these models also predict trends at odds with the observed ones for s-process elements, especially for Zr and Ba.
    Therefore, the analysis rejects the i-process in AGB stars as an important contributor in n-capture elements evolution. However, we cannot exclude smaller contributions due to the large uncertainties affecting such nucleosynthetic calculations.
    
\end{itemize}

The chemical modelling of objects such as the SMC will certainly be an object of interest in the next years, thanks to the advent of next generation multi-object spectrographs (MOONS, 4MOST) and related surveys \citep{Cioni19,Gonzalez20}. 
In this regard, this work can already give important information on the chemical evolution of this object, providing a test-bed in the context of the next-to-happen data revolution.

\begin{acknowledgements}
The authors thank the anonymous referee for the careful reading of the manuscript and the suggestions provided. 
MP, AM, and DR acknowledge financial support from the project "LEGO – Reconstructing the building blocks of the Galaxy by chemical tagging" granted by the Italian MUR through the contract PRIN2022LLP8TK\_001 (PI A.~Mucciarelli). 
MP also acknowledges support from HORIZON-INFRA-2024-DEV-01-01 – Research Infrastructure Concept Development, through the project WST: The Wide-Field Spectroscopic Telescope (Grant No. 101183153).  
FM thanks I.N.A.F. for the 1.05.24.07.02 Mini Grant - LEGARE "Linking the chemical Evolution of Galactic discs AcRoss diversE scales: from the thin disc to the nuclear stellar disc" (PI E. Spitoni).
FM  also  thanks I.N.A.F. for the 1.05.12.06.05 Theory Grant - Galactic archaeology with radioactive and stable nuclei.
FM  also acknowledges support from Project PRIN MUR 2022 (code 2022ARWP9C) "Early Formation and Evolution of Bulge and HalO (EFEBHO)" (PI: M. Marconi).
\end{acknowledgements}

%
   \bibliographystyle{aa} 
   \bibliography{biblio} 
%


\appendix

\section{Sample comparison with APOGEE stars}
\label{a:data_comparison}

In the main text, we compare the model predictions for [X/Fe] vs. [Fe/H] with the dataset presented in \citetalias{Mucciarelli23} and \citetalias{Anoardo25}, for which we can count on an extensive chemical inventory for n-capture elements (Zr, Ba, La, Eu).
This is not available for APOGEE data; however, it is worth comparing APOGEE stars  with the ones shown in the main text for lighter elements (up to the Fe-peak). In this way, it is possible to rule out or not the presence of spurious pattern \citetalias{Mucciarelli23} that can affect our model calibration.

\begin{figure*}
    \centering
    \includegraphics[width=0.925\linewidth]{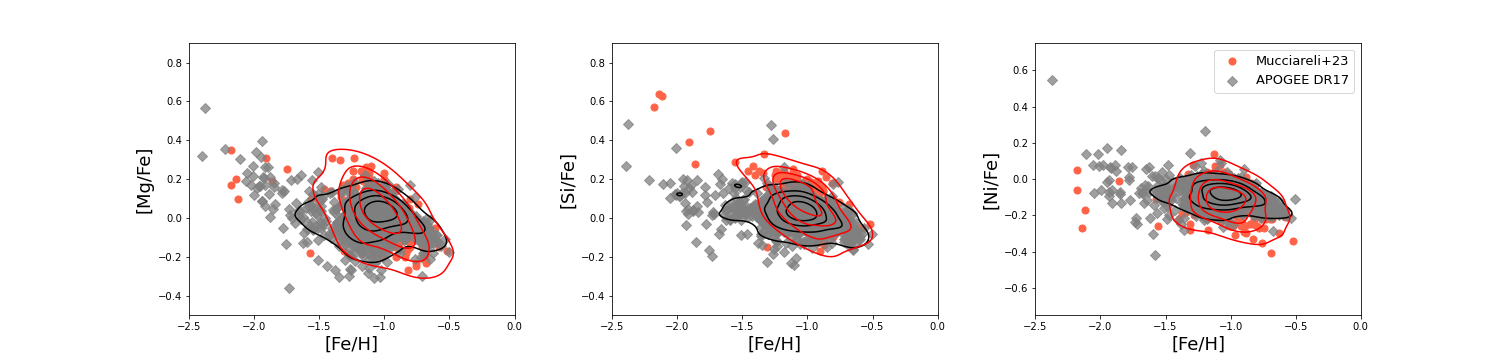}
    \caption{Comparison between [X/Fe] vs. [Fe/H] observed abundance patterns for light elements in the SMC. Orange dots are from \citet{Mucciarelli23}, whereas grey diamonds are APOGEE DR17 data (\citealt{Hasselquist21}). 
    Orange and black contour lines represent density lines of the observed stellar distributions in \citet{Mucciarelli23} and \citet{Hasselquist21}, respectively.}
    \label{fig:comparison_APOGEE}
\end{figure*}

The comparison between the [Mg,Si,Ni/Fe] vs [Fe/H] patterns as in \citetalias{Mucciarelli23} and in APOGEE DR17 \citep{Hasselquist21} is shown in Fig. \ref{fig:comparison_APOGEE}.
The abundance patterns show very similar behaviour in the observed metallicity ranges, without any significant offset. Indeed, data density contours of the tow datasets significantly superimpose in the three analysed elements, ruling out significant discrepancies.

\section{Model parameter recalibration according to IMF}
\label{a:IMF_calibration}

As mentioned in \ref{ss:IMF}, changing the IMF in the GCE model has consequences not only on the production of n-capture elements, but also on the evolution of other metals. 
Therefore, the models adopting a different IMF from the canonical \citet{Kroupa01} are re-calibrated in the SFE parameter $\nu$ to reproduce the SMC observables other than n-capture elements. 
In this way, we still rely on models with chemical evolution histories in agreement with the observations displayed in Section \ref{s:results}, which is key to perform our analysis on the origin of n-capture elements in Section \ref{s:ncapture_results}.\\

\begin{figure*}
    \centering
    \includegraphics[width=0.925\linewidth]{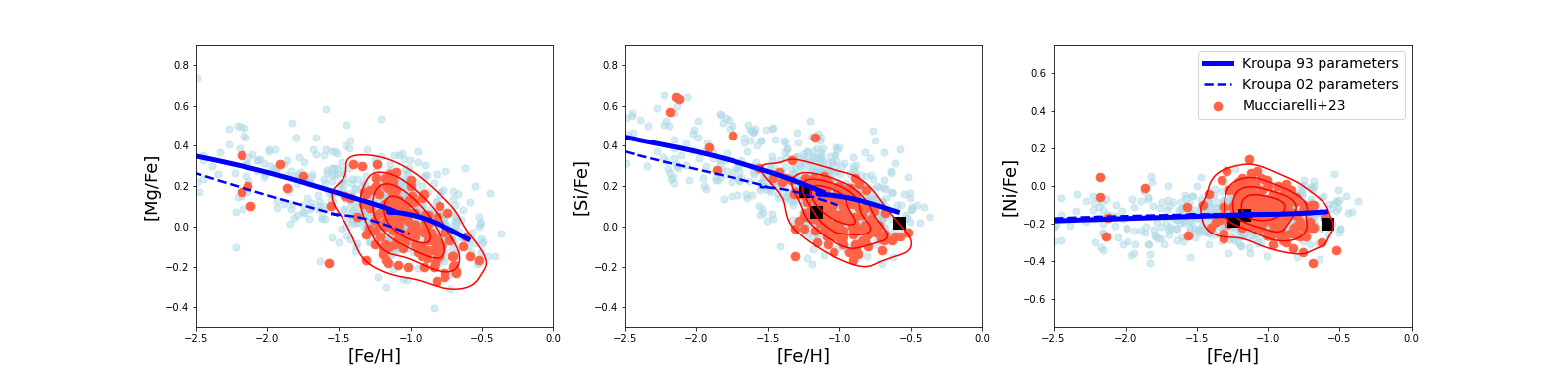}
    \includegraphics[width=0.925\linewidth]{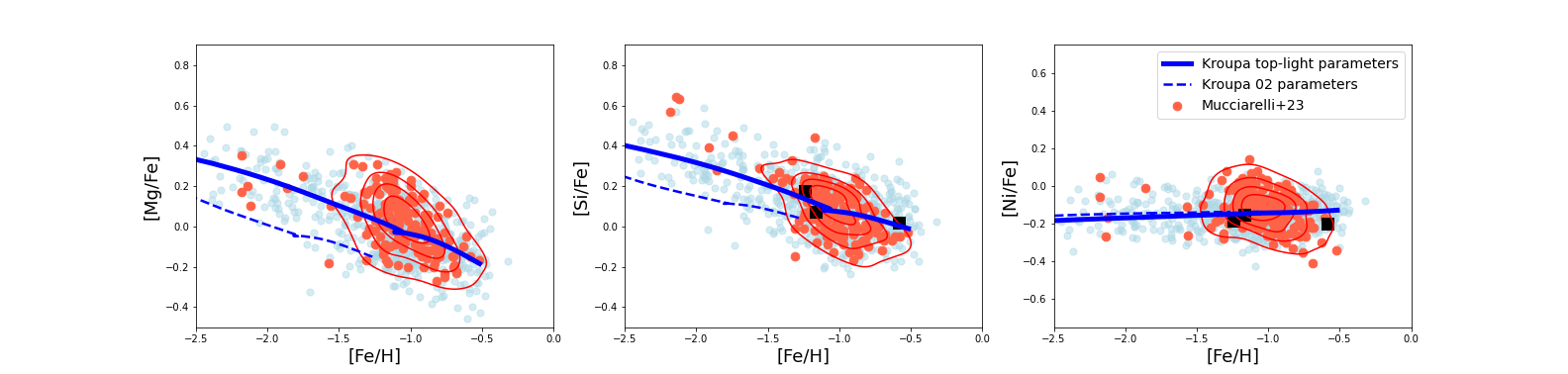}
    \caption{[X/Fe] vs. [Fe/H] for light elements in the SMC adopting different IMFs. Abundance patterns are shown for models adopting a \citet{Kroupa93} IMF (top panels) and a modified \citet{Kroupa93} IMF with steeper high-mass end slope (top-light IMF, bottom panels).
    Solid lines are genuine chemical evolution tracks for models with the re-calibrated SFEs adopted for the particular IMF setup, while dashed lines are models with the same SFEs used for the reference \citet{Kroupa01} IMF. Cyan shaded regions are predictions for the model with the re-calibrated SFEs adopted for the IMF setup, accounting for observational uncertainties ('synthetic model').
    Data are as in Fig. \ref{fig:XFe_FeH}.}
    \label{fig:IMF_K93_toplight}
\end{figure*}

In Fig. \ref{fig:IMF_K93_toplight}, we show the comparison between the observed and predicted [X/Fe] vs. [Fe/H] for light ($\alpha$ and Fe-peak) elements for both the case of the \citet{Kroupa93} IMF (top panels) and the adopted top-light IMF (\citealt{Kroupa93} with high mass slope $x=-2$, bottom panels). In particular, solid lines and cyan shaded regions display the results for models adopting the re-calibrated SFEs for the specific IMF setup. Both top and bottom panels show good agreement between data and model predictions, similar to what happens to the model with the reference \citet{Kroupa01} IMF (see Fig. \ref{fig:XFe_FeH}).

To achieve such agreement, in the model with the \citet{Kroupa93} IMF we adopt a SFE of $\nu=0.04$ Gyr$^{-1}$ up to $t=$ 13 Gyr, and $\nu=0.1$ Gyr$^{-1}$ in the last Gyr, whereas for the model with top-light IMF we assume a SFE of $\nu=0.08$ Gyr$^{-1}$ up to $t=$ 13 Gyr, and $\nu=0.2$ Gyr$^{-1}$ in the last Gyr. 
We remind that the increase in SFE in the late stages of the model evolution can be explained by the large and rapid increase in SFR found by \citetalias{Rubele18} SFH in the last Gyr, which will cause a gas dilution effect too strong without increasing the SFE parameter (see also Section \ref{s:results}).

On the other hand, we note that both the models with \citet{Kroupa93} and top-light IMF display larger SFE relative to the one adopting the reference \citet{Kroupa01} IMF. This is due to the lower contribution of massive stars to chemical enrichment, which, without an increase in the SFE, causes a i) a stark decrease in metallicity at a given age and ii) a decrease in the level of the [$\alpha$/Fe] enrichment. This is clearly shown in Fig. \ref{fig:IMF_K93_toplight}, where the dashed lines represent the predictions obtained for models with different IMF while using the SFEs assumed for the reference \citet{Kroupa01} IMF.

\section{Additional model runs}
\label{a:addition_model}

In the following, we provide figures for additional model tracks and comparison with the samples of \citetalias{Mucciarelli23} and \citetalias{Anoardo25} that are not shown in the main text. \\

\begin{figure*}
    \centering
    \includegraphics[width=1.0\linewidth]{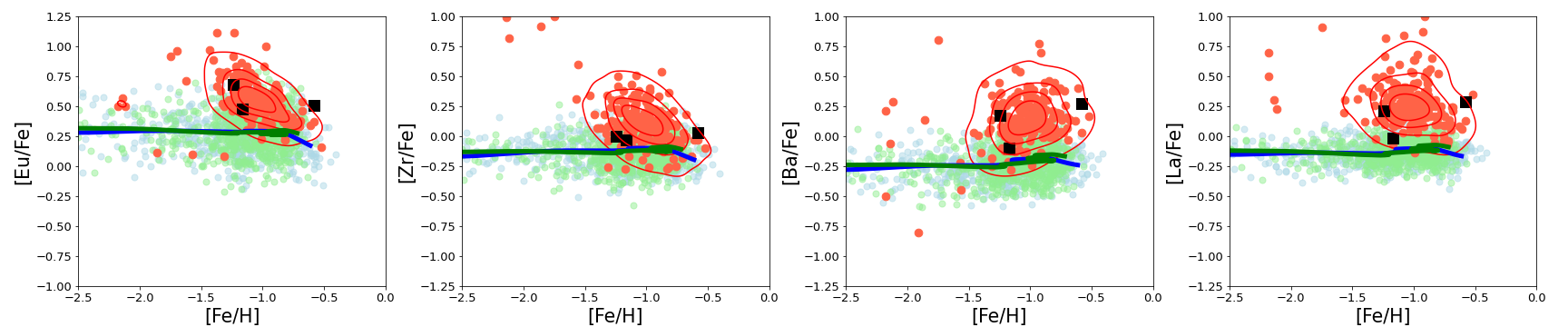}
    \caption{[X/Fe] vs. [Fe/H] for n-capture elements for models with increased MNS r-process production at low-metallicity (see Eq. \eqref{eq:alpha_incr}). Solid lines are genuine chemical evolution tracks for the reference models adopting \citet{Rubele18} SFH (blue) and \citet{Massana22} SFH (green), whereas shaded cyan and light-green regions are associated model predictions account for observational uncertainties ('synthetic models'). Data are as in Fig. \ref{fig:ncFe_std}.}
    \label{fig:ncFe_moreMNS_compareSFH}
\end{figure*}

First, in Fig. \ref{fig:ncFe_moreMNS_compareSFH} we show the predictions for n-capture elements (Eu, Zr, Ba, La) abundance ratios for models with increased MNS r-process production as suggested by \citet{Palla25}. At variance with Fig. \ref{fig:ncFe_moreMNS}, where only the outcome by the model with \citetalias{Rubele18} SFH is shown, here we also plot the results of the models adopting \citetalias{Massana22} SFH. 
Despite Fig. \ref{fig:ncFe_std} showing already remarkable similarities in the n-capture elements predicted patterns by the models employing the different SFHs, it is important to test the outcome by the two different models in light of the changes in the prescriptions for r-process production. Indeed, introducing a metallicity threshold in the r-process producers may cause additional important changes in the [X/Fe] vs. [Fe/H] patterns for n-capture elements produced by the different SF and chemical enrichment histories.
However, Fig. \ref{fig:ncFe_moreMNS_compareSFH} excludes this possibility: the predicted abundance patterns for the models with different SFH remain very similar throughout the galaxy evolution.

\begin{figure*}
    \centering
    \includegraphics[width=0.925\linewidth]{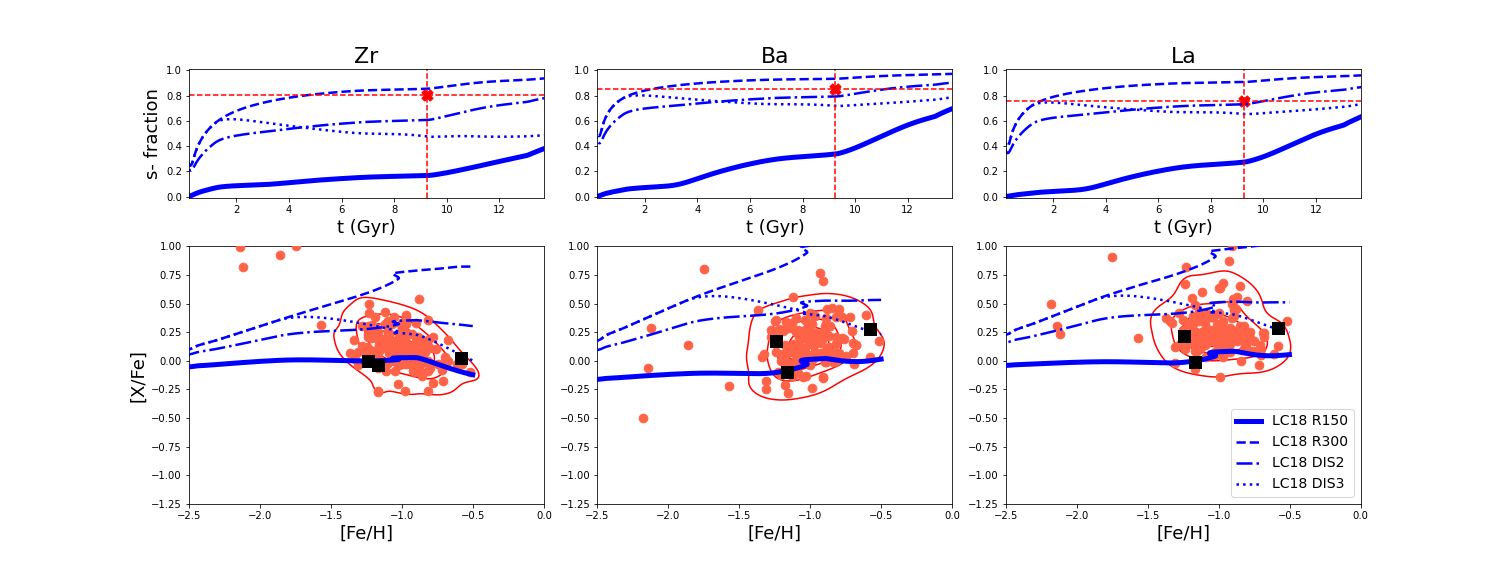}
    \caption{Mass fraction produced by the s-process channel of production (top panels) and [X/Fe] vs. [Fe/H] (bottom panels) for Zr (left panels), Ba (central panels) and La (right panels) for models with different IRV distribution and \citet{Kroupa93} IMF. Results are shown for models adopting the reference IRV distribution (R150, thick solid lines), R300 distribution (thin dashed), DIS2 distribution (thin dash-dotted) and DIS3 distribution (thin dotted). Data are as in Fig. \ref{fig:ncEu_sfrac}.}
    \label{fig:IMF_IRV}
\end{figure*}

Fig. \ref{fig:IMF_IRV} displays the [s-process/Fe] vs. [Fe/H] (top panels) and s-process fraction vs. evolutionary time (bottom panels) for models adopting IMF and IRV different from the reference ones.
For the sake of Figure clarity, we show the predictions only for the models adopting a \citet{Kroupa93} IMF and different IRV setup compared with the reference IRV adopting the same IMF.
As the Figure shows, variations in IRV distribution reflect the changes seen in Fig. \ref{fig:different_IRVs}. Similar variations are also observed for the top-light IMF as defined in Section \ref{ss:IMF}.

\begin{figure*}
    \centering
    \includegraphics[width=0.99\linewidth]{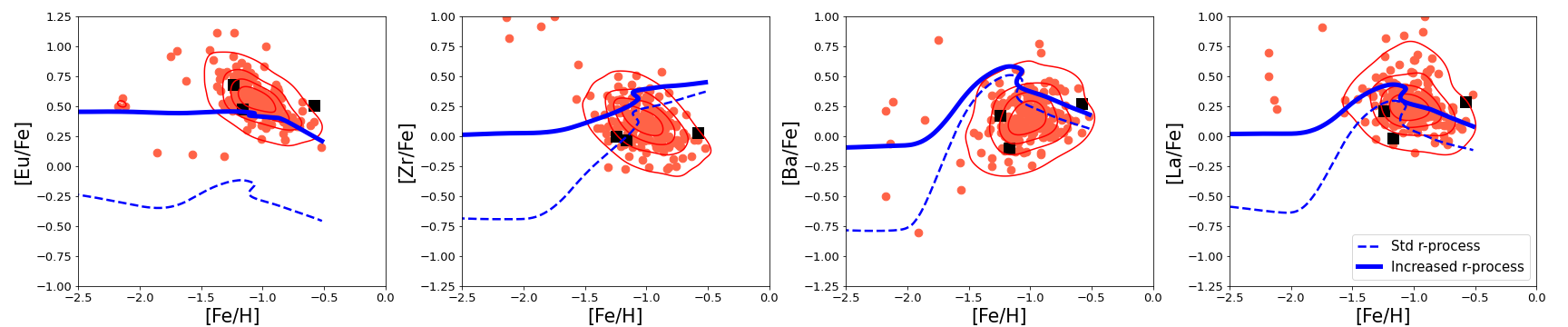}
    \caption{[X/Fe] vs. [Fe/H] for n-capture elements for models with \citet{Rubele18} SFH, different rate of MNS r-process production at low-metallicity, low-mass AGB yields by \citet{Choplin22,Choplin24}, adopting the top-light IMFs defined in \ref{ss:IMF}. Blue lines represent models adopting standard (thin dashed) and increased MNS r-process production at low-metallicity (thick solid). Data are as in Fig. \ref{fig:ncFe_std}.}
    \label{fig:ncFe_iproc_MNS}
\end{figure*}

In Fig. \ref{fig:ncFe_iproc_MNS}, we display instead the abundance patterns for n-capture elements either with a standard r-process enrichment setup (as shown in the main text) or with increased r-process production by MNS at low-metallicity, when adopting the yields by \citet{Choplin22,Choplin24} for low-mass stars. For the sake of Figure clarity and avoid overcrowding, the predictions are shown for the top-light IMF defined in Section \ref{ss:IMF}.
The panels show that even when an additional enrichment component is considered for n-capture elements, the predicted [X/Fe] patterns still struggle to be reproduced when current state-of-the-art stellar yields including i-process production are considered. This is especially evident when looking at Zr and Ba, where the predicted rising and decreasing trends are at odds with observed ones (decreasing and rising, respectively).

\end{document}